\documentclass[10pt,twocolumn]{article}
\usepackage{xcolor}
\usepackage{amsmath}
\usepackage{graphicx}
\usepackage{authblk}
\usepackage{amssymb}
\usepackage{hyperref}
\usepackage{cite}

\usepackage[top=2cm, bottom=2cm, left=1.5cm, right=1.5cm]{geometry}

\newcommand{\be}{\begin{equation}}
\newcommand{\ee}{\end{equation}}
\newcommand{\bg}{\begin{equation}}
\newcommand{\eg}{\end{equation}}
\newcommand{\bdm}{\begin{displaymath}}
\newcommand{\edm}{\end{displaymath}}
\newcommand{\bea}{\begin{eqnarray}}
\newcommand{\eea}{\end{eqnarray}}
\newcommand{\beas}{\begin{eqnarray*}}
\newcommand{\eeas}{\end{eqnarray*}}
\newcommand{\ba}{\begin{array}}
\newcommand{\ea}{\end{array}}
\newcommand{\nn}{\nonumber}

\newcommand{\bfg}{\begin{figure}}
\newcommand{\efg}{\end{figure}}

\newcommand{\fr}{\frac}

\newcommand{\mb}{\mbox}

\newtheorem{lm}{Lemma}
\newtheorem{cl}{Corollary}
\newtheorem{df}{Definition}

\newcommand{\blm}{\begin{lm}}
\newcommand{\elm}{\end{lm}}
\newcommand{\bcl}{\begin{cl}}
\newcommand{\ecl}{\end{cl}}
\newcommand{\bdf}{\begin{df}}
\newcommand{\edf}{\end{df}}
\newcommand{\brk}{\begin{rm}}
\newcommand{\erk}{\end{rm}}
\newcommand{\hst}{\hspace*}

\newcommand{\lb}{\label}

\newcommand{\om}{\omega}
\newcommand{\Om}{\Omega}

\newcommand{\veps}{\varepsilon}

\newcommand{\ld}{\lambda}

\newcommand{\rf}{\ref}

\title{Intuitive understanding of extinction of small particles in absorbing and active host media within the MLWA}

\author[1]{Anton D. Utyushev}
\author[2,3]{Vadim I. Zakomirnyi}
\author[1]{Alexey A. Shcherbakov}
\author[4]{Ilia L. Rasskazov}
\author[5]{Alexander Moroz}

\affil[1]{School of Physics and Engineering, ITMO University, 197101, Saint-Petersburg, Russia}
\affil[2]{Beckman Institute for Advanced Science and Technology, University of Illinois at Urbana-Champaign, Urbana, IL 61801, USA}
\affil[3]{Lumière, nanomatériaux, nanotechnologies (L2n), UMR CNRS 7076, Université de Technologie de Troyes, Troyes 10004, France}
\affil[4]{SunDensity Inc., Rochester, NY 14604, USA}
\affil[5]{Wave-scattering.com}

\date{}

\begin{document}

\maketitle

\begin{abstract} 
In an absorbing or an active host medium characterized by a complex refractive index $n_2=n_2'+{\rm i}n_2''$, our previously developed modified dipole long-wave approximation (MLWA) is shown to essentially overly with the exact Mie theory results for localized surface plasmon resonance of spherical nanoparticles with radius $a\lesssim 25$~nm ($a\lesssim 20$~nm) in the case of Ag and Au (Al and Mg) nanoparticles.
The agreement for Au and Ag (Al and Mg) nanoparticles, slightly better in the case of Au than Ag, continues to be acceptable up to $a\sim 50$~nm ($a\sim 40$~nm), and can be used, at least qualitatively, up to $a\sim 70$~nm ($a\sim 50$~nm) correspondingly.
A first order analytic perturbation theory (PT) in a normalized extinction coefficient, $\bar\kappa=n_2''/n_2'$, around a nonabsorbing host is developed within the dipole MLWA and its properties are investigated.
It is shown that, in a suitable parameter range, the PT can reliably isolate and capture the effect of host absorption or host gain on the overall extinction efficiency of various plasmonic nanoparticles. 
\end{abstract}

%%%%%%%%%%%%%%%%%%%%%%%%%%  body  %%%%%%%%%%%%%%%%%%%%%%%%%%

\section{Introduction}
%%%%%%%%%%%%%%%%%%%%%%
Electromagnetic scattering in an absorbing host characterized by a complex refractive index $n_2=n_2'+{\rm i}n_2''$ (where $n_2'$ and $n_2''$ are real) has more than fifty years old history. 
The traditional scattering theory neglects the host dissipation and gain~\cite{Newton1982,Bohren1998}, because those cases imply either vanishing or infinite scattering wave at the spatial infinity.
Once wave number $k=k'+{\rm i}k''=2\pi n_2/\ld_0$, where $\ld_0$ is the {\em vacuum} wavelength, is a complex number, i.e. $k''\ne 0$, conventional expressions for cross sections cannot be straightforwardly extended for $k''\ne 0$, because the expressions yield cross sections as {\em complex} quantities.  
Not surprisingly, the history of scattering in an absorbing host is filled in with a number of controversies~\cite{Fuchs1968,Mundy1974,Chylek1979,Bohren1979,Lebedev1999,Sudiarta2001,Yang2002,Fu2006,Mishchenko2007,Mishchenko2017,Mishchenko2018,Mishchenko2019,Peck2019,Mishchenko2019b,Khlebtsov2021,Dong2021,Zhang2022b}. 
This is probably why in classical textbooks it is only fleetingly mentioned in Section~12.1.3 of ref.~\citenum{Bohren1998}.
Already the definition of an incident intensity is not straightforward, as the field incident on a particle is different at different points on the particle as a result of an absorbing medium~\cite{Mundy1974,Chylek1979}.

In contrast to the conventional case of $k''= 0$, two sets of cross sections are commonly used: {\em inherent} and {\em apparent}.
The former is obtained by performing surface integrals of corresponding Poynting vectors over the particle surface.
The approach was developed a long time ago~\cite{Mundy1974,Chylek1979}, but currently accepted expressions for the {\em inherent} cross-sections were not presented until 1999~\cite{Lebedev1999,Sudiarta2001,Yang2002,Fu2006}.
The focus of present work will be on the apparent extinction cross-section $C_{\rm ext}$, which is operationally defined in the far field (see Section~\ref{sc:aecs} below).
The history of {\em apparent} cross sections began with the optical theorem of Bohren and Gilra~\cite{Bohren1979}.
Here, too, it took a long time, until 2007, to arrive at definitive expressions for the apparent extinction cross-section cross section, $C_{\rm ext}$~\cite{Mishchenko2007,Mishchenko2017,Mishchenko2018,Mishchenko2019}.
A curious feature of the apparent extinction cross-section in an absorbing host is that it can be {\em negative}~\cite{Mishchenko2019a}, which is neither an artifact of numerical simulations nor it violates any physical law.
The apparent extinction cross-section quantifies the difference in the readings of a forward-scattering detector taken with and without the particle.
If the surrounding medium is absorbing, the presence of the particle can in fact make the detector signal stronger, thereby implying a negative extinction cross-section.
There is no violation of the energy conservation law, since in this case the extinction cross-section is not used to quantify the energy budget of a finite volume encompassing the particle~\cite{Mishchenko2019a}. 
Another curiosity of apparent cross-sections in an absorbing host is that an intrinsic definition of an apparent absorption cross-sections, $C_{\rm abs}$, is still missing.
The difficulty lies in that the very presence of a particle necessarily modifies the near field around the particle.
The latter may be the cause of an additional absorption in the host medium {\em outside} the particle compared to what is happening in the absence of the particle~\cite{Sudiarta2001,Yang2002}.
The critical point is that because the field is disturbed by the particle, additional absorption may be realized in the medium external to the particle~\cite{Videen2003}. 
Consequently, unlike $C_{\rm abs}$ and $C_{\rm sca}$, only apparent extinction cross-section $C_{\rm ext}$ can be consistently defined (see Section~\ref{sc:aecs} below).

The focus of present work is on the apparent extinction cross-section $C_{\rm ext}$ in the so-called {\em modified long-wave approximation} (MLWA)~\cite{Meier1983,Zeman1984,Zeman1987,Kelly2003,Kuwata2003,Moroz2009,Zoric2011,Massa2013,LeRu2013,Schebarchov2013,Januar2020,Rasskazov2020a,Rasskazov2021,Khlebtsov2021,Zhang2022b}.
Firstly, it will be shown that the apparent extinction cross-section conforms better to physical intuition than the intrinsic extinction cross-section. 
Secondly, the MLWA~\cite{Meier1983,Zeman1984,Zeman1987,Kelly2003,Kuwata2003,Moroz2009,Zoric2011,Massa2013,LeRu2013,Schebarchov2013,Januar2020,Rasskazov2020a,Rasskazov2021,Khlebtsov2021,Zhang2022b}, which can be viewed as a next-order approximation beyond the Rayleigh limit, is known to overcome a number of severe deficiencies of the Rayleigh limit (see Eq.~(\rf{mlwaff}) below) and, at least for nonabsorbing hosts, be surprisingly precise~\cite{Moroz2009,Schebarchov2013,Rasskazov2020a,Rasskazov2021,Khlebtsov2021,Zhang2022b}. 

\begin{figure}[!t]
\centering
\includegraphics[width=3in]{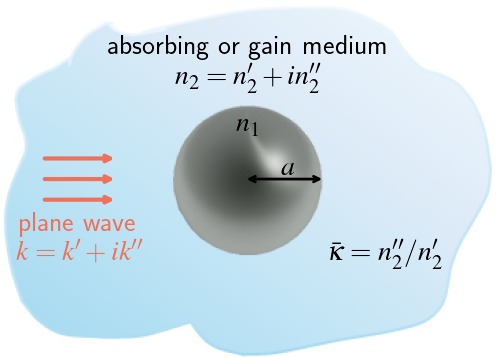}
\caption{Schematic representation of the problem under consideration. A spherical particle with radius $a$ and refractive index $n_1$ is embedded in an absorbing or gain medium with $n_2=n_2'+{\rm i}n_2''$, where $n_2'$ and $n_2''$ are real. 
Imaginary refractive index of the host $n_2''>0$ for an absorbing medium and $n_2''<0$ for a gain medium, and $n_2''=0$ for a transparent medium.
Normalized extinction coefficient $\bar\kappa=n_2''/n_2'$ is introduced to quantify the magnitude of the host dissipation or gain.}
\label{fig:mlwa_scheme}
\end{figure}

The outline of our contribution is as follows.
Section~\ref{sc:th} first  recalls in its Subsection~\ref{sc:aecs} the expression for distance-independent apparent extinction cross section $C_{\rm ext}$ in the framework of the Lorenz–Mie theory~\cite{Mishchenko2018}.
In Subsection~\ref{sc:mlwa} our earlier developed MLWA~\cite{Rasskazov2020a,Rasskazov2021} is summarized.
In Subsection~\ref{sc:mlwapt}, still within the framework of MLWA, an analytic first-order perturbation theory (PT) in a normalized extinction coefficient, $\bar\kappa=n_2''/n_2'$, representing the magnitude of the host dissipation, is developed.
A motivation for developing the above PT is to isolate and capture the effect of the host absorption or host gain on the overall extinction, which is not possible within the MLWA (the latter captures only overall extinction).
In Section~\ref{sc:res}, performance of the above approximations in describing localized surface plasmon resonances of spherical plasmonic nanoparticles in different absorbing and active hosts is investigated.
Discussion of some of the observed features is provided in Section~\ref{sc:disc}.
We then conclude with Section~\ref{sc:conc}.

\section{Theory}
\lb{sc:th}
%%%%%%%%%%%%%%%%%%%
\subsection{Apparent extinction cross section in the framework of the Lorenz–Mie theory}
\lb{sc:aecs}
%%%%%%%%%%%%%%%%%%%%%%%%%
The specific expression for distance-independent apparent extinction cross section $C_{\rm ext}$ in the framework of the Lorenz–Mie theory is~\cite{Mishchenko2018} is given as an infinite sum over different
multipole orders $\ell\ge 1$ (see Fig.~\ref{fig:mlwa_scheme}),
\bg
C_{\rm ext}= - \fr{2\pi}{k'}\, \mb{Re }\left\{ \sum_{\ell=1}^\infty \fr{1}{k}\, (2\ell+1)(T_{E\ell}+T_{M\ell})\right\}.
\lb{Cext}
\eg
In the optical convention, $T_{p\ell}$ correspond to the familiar Mie's expansion coefficients $a_\ell$ and $b_\ell$ (Eqs.~(4.53) of ref.~\citenum{Bohren1998}), i.e. $T_{E\ell} = -a_\ell$ for electric, or transverse magnetic (TM), polarization, and $T_{M\ell} = -b_\ell$ for magnetic, or transverse electric (TE), polarization. 
The physical, or operational, meaning of the distance-independent apparent $C_{\rm ext}$ is that it determines the reading of a polarization-sensitive well-collimated radiometer (WCR) at a sufficiently large distance $r$ from the particle~\cite{Mishchenko2018}, 
\bg
\mb{WCR signal} \propto \exp(-2k'' r) (\Om-C_{\rm ext}) I_{\rm inc},
\nonumber
\eg
where $\Om$ is the area of the objective lens of the WCR, and $I_{\rm inc}$ is the intensity of the incident homogeneous (uniform) plane wave at the center of the particle. 
Importantly, the distance-independent extinction cross-section {\em cannot} be introduced in the context of evaluating the energy budget of an arbitrarily shaped volume containing the scattering particle~\cite{Mishchenko2018}.

\subsection{The dipole MLWA}
\lb{sc:mlwa}
%%%%%%%%%%%%%%%%%%%%%%%%%%%%%%%%
\begin{figure}[!ht]
\centering
\includegraphics[width=2.3in]{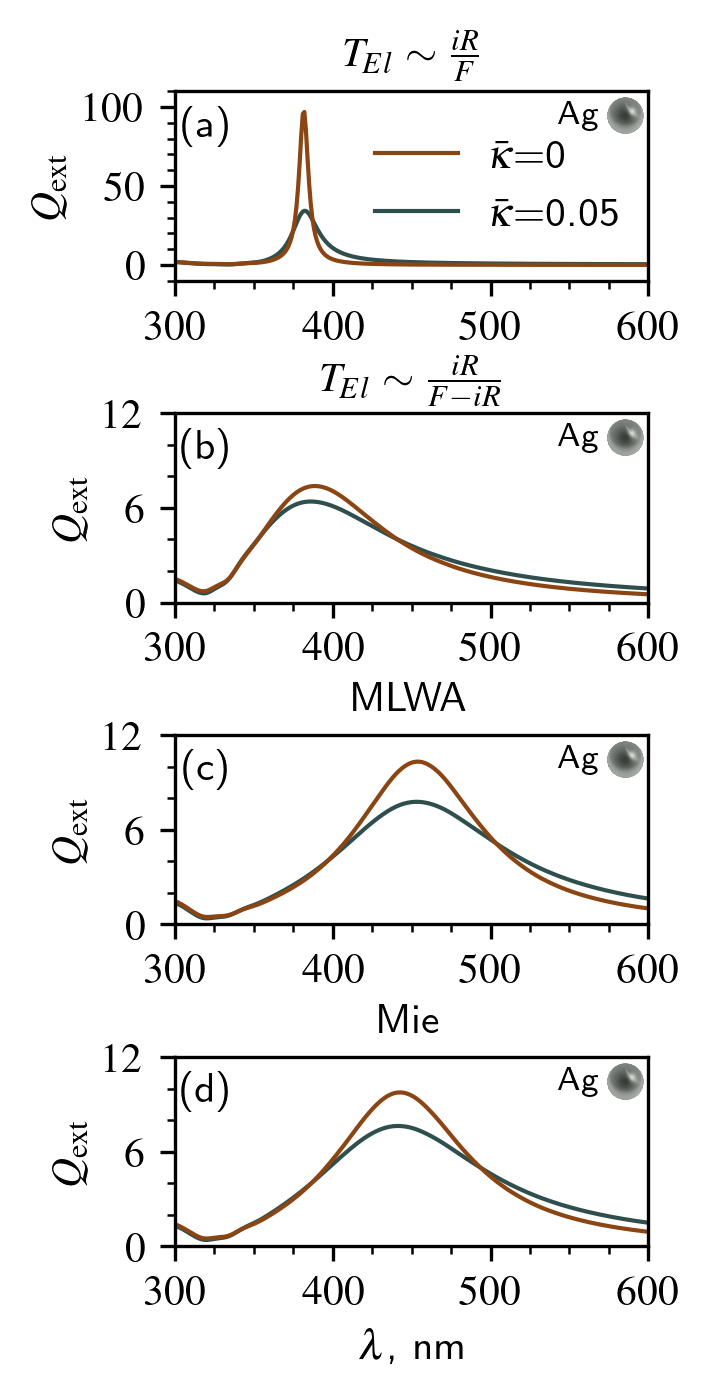}
\caption{Demonstration of different physical mechanisms described within MLWA on the extinction spectra of Ag sphere with $a=40$~nm in the host with $n_2'=1.33$.
Here $\bar \kappa=0$ for a transparent host and $\bar\kappa=0.05$ in case of a dissipative host.
The refractive indices of Ag were taken from ref.~\citenum{McPeak2015}.
(a) Quasi-static Rayleigh approximation;
(b) radiative correction taking the effect of retardation with respect to the incident field;
(c) the MLWA including additionally the dynamic depolarization term.
Obviously all the correction terms of the dipole MLWA given by Eq.~\eqref{mlwa1} are  necessary to achieve a fairly accurate approximation to the exact Mie theory shown in panel (d) for a comparison.
All the above approximations, but the exact Mie theory, assume a constant field inside the sphere.}
\label{fig:mlwa_add}
\end{figure}
%%%%%%%%%%%%%%%%%%%%%%%%%%%%%%%%%%%%%%%%%%%%%%%%%%%%%%%%%%%%%%%%%%
The MLWA is a rational approximation to the Mie coefficients in terms of a fraction of simple polynomials in size parameter $x$ that in a concise way combines three different elementary terms, involving {\em size-independent} quasi-static Fr\"ohlich term, 
\bg
F_{p\ell} := \upsilon + \fr{\ell+1}{\ell},
\nonumber  %\lb{FT}
\eg
the {\em dynamic depolarization} ($D_{p\ell} \sim x^2$) and the {\em radiative reaction} ($R_{p\ell}\sim x^{2\ell +1}$ for $p=E$ and $R_{p\ell}\sim x^{2\ell +3}$ for $p=M$) in the functional form~\cite{Meier1983,Zeman1984,Zeman1987,Kelly2003,Kuwata2003,Moroz2009,Zoric2011,Massa2013,LeRu2013,Schebarchov2013,Januar2020,Rasskazov2020a,Rasskazov2021,Khlebtsov2021,Zhang2022b}
\bg
T_{p\ell} = \fr{i R_{p\ell}(x)} {F_{p\ell} + D_{p\ell}(x) -i R_{p\ell}(x)}\cdot
 \lb{mlwaff}
\eg
Assuming nonmagnetic media, one has $\upsilon=\mu_1/\mu_2=1$ for magnetic (TE) polarization ($p=M$), and $\upsilon=\veps:=\veps_1/\veps_2$ for electric (TM) polarization ($p=E$), where the subscript 1 (2) identifies the relevant quantities of a sphere (host).

The respective terms $F_{p\ell}$, $D_{p\ell}$, $R_{p\ell}$ in the functional form of Eq.~\eqref{mlwaff} have well-defined physical origin and meaning (see Fig.~\ref{fig:mlwa_add}). This allows for an intuitive understanding of scattering from small particles which, as it will be shown, allows for rather reliable substitution of infinite series expansion in terms of Bessel functions of the conventional Mie solution by a single dipole term.
The functional form of Eq.~\eqref{mlwaff} makes it transparent that the usual Rayleigh limit, which amounts to setting $D_{p\ell}(x)=R_{p\ell}(x)\equiv 0$ in the denominator for $\ell=1$ and $p=E$, is recovered for $x, x_s\ll 1$.
Here $x$ is in general complex size parameter, $x=2\pi a/\lambda$, with $\lambda$ being the wavelength in the host medium, whereas $x_s=2\pi n_2 a/n_1\lambda=x/\sqrt{\veps}$.
The vanishing of the size-independent $F$ in the denominator yields the usual quasi-static Fr\"ohlich condition which determines the quasi-static frequencies $\om_{0\ell}$ for the occurrence of a localized surface plasmon resonance (LSPR).
In what follows, we shall focus on the dipole MLWA which yields~\cite{Rasskazov2020a,Rasskazov2021}
\bg
T_{E1} \approx \frac{2i (\veps-1)x^3/3} {\veps +2 -3(\veps-2)x^2/5 -2i (\veps-1)x^3/3},
\lb{mlwa1}
\eg
where $\veps=\veps_1/\veps_2$ is the relative dielectric function. 
An exceptional feature of the MLWA dipole contribution is that one can determine analytically an exact position of the complex pole of $T_{E1}$ (the Mie coefficient $-a_1$), and hence the dipolar LSPR position, at
\bea
\veps_{E1} = -2 \times \fr{1+3x^2/5+ix^3/3}{1-3x^2/5-2ix^3/3},
\lb{vepsp}
\eea
which corrects formula Eq.~(6) of ref.~\citenum{Zhang2022b}.
The proof of that $\veps_{E1}$ yields complex zero of the denominator $D$ of $a_1$ is relegated to Section~S3 in Supplement 1.
In the limit of small $x$, one can expand the denominator of $\veps_{E1}$ as $\sim 1+3x^2/5+2ix^3/3$, whereby Eq.~\eqref{vepsp} reduces to the familiar classical Bohren and Huffman result (cf. Eq.~(12.13) of ref.~\citenum{Bohren1998}):
\begin{equation}
\veps_{E1} \sim \veps_{\rm BH} \approx -2-\frac{12 x^2}{5}\qquad (|x|\ll 1).
\label{vepsz}
\end{equation}

%%%%%%%%%%%%%%%%%%%%%%
\begin{figure*}[!ht]
\centering
\includegraphics[width=\linewidth]{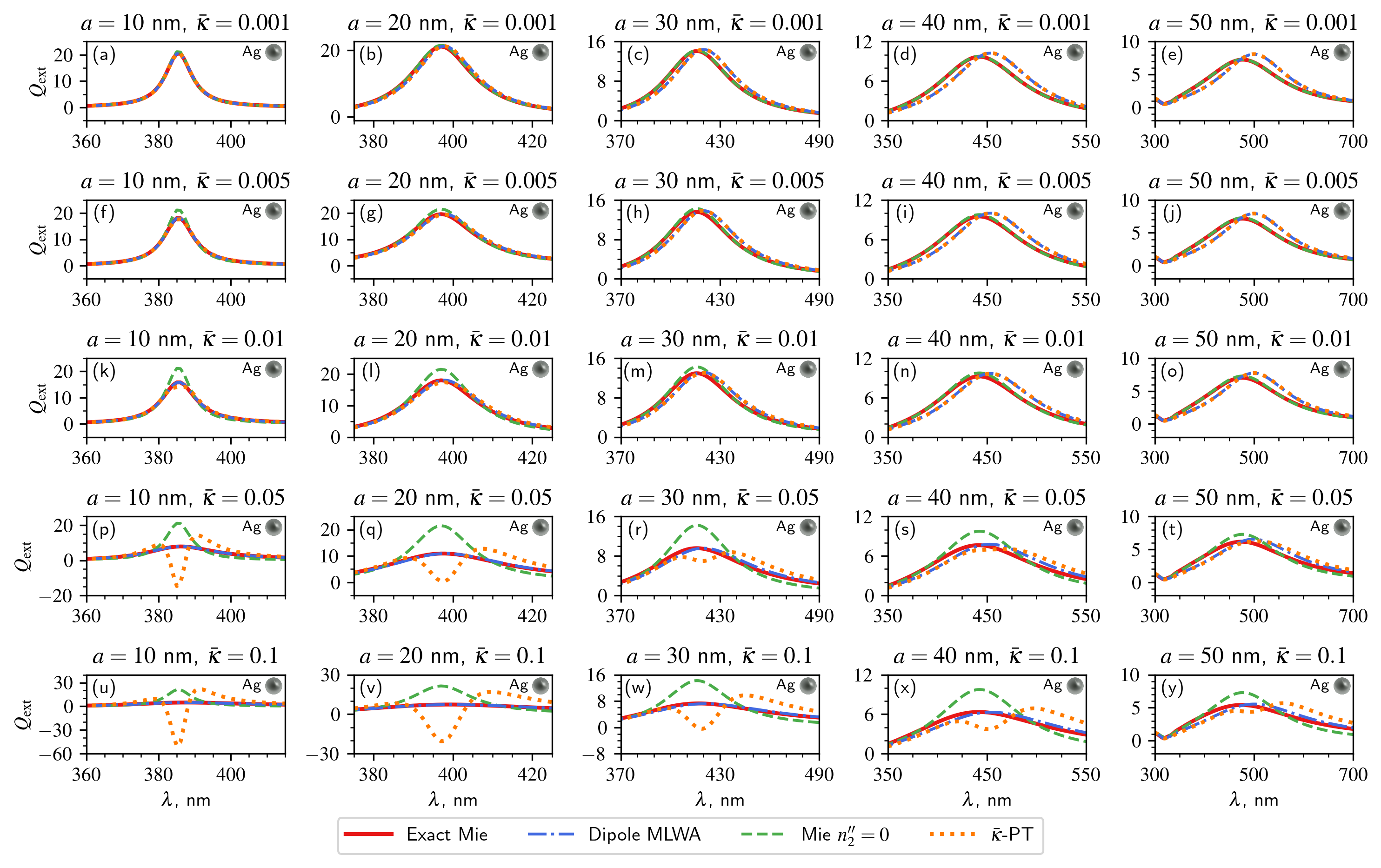}
\caption{Extinction spectra, $Q_{\rm ext}$, for individual spherical Ag nanoparticles with radii $a=10$, $20$, $30$, $40$, $50$~nm in water-like host ($n_2'=1.33$) with $\bar\kappa = 0.001$, $0.005$, $0.01$, $0.05$, $0.1$, as shown in the legend for each plot.
Spectra were calculated by Eq.~(\ref{Cext}) of the exact Mie theory (solid red line), the dipole MLWA by Eq.~(\ref{mlwa1}) (blue dot-dashed line), the Mie theory in nonabsorbing host ($n_2''=0$;
dashed green line), and the first order $\bar\kappa$-PT of our Eq.~(\ref{Cextl1}) (dotted orange line).
Because the Mie theory in the nonabsorbing host is independent of $\bar\kappa$, its plots are identical at each given column.
The refractive indices of Ag were taken from ref.~\citenum{McPeak2015}.
Note a broadening and suppression of plasmon resonances with increasing host absorption.}
\label{fig:mlwa_tau}
\end{figure*}
%%%%%%%%%%%%%%%%%%

\subsection{Perturbation theory in a normalized extinction coefficient within the MLWA}
\lb{sc:mlwapt}
%%%%%%%%%%%%%%%%%%%%%%%%%%%%%%%%%%%%%%%%%%%%%%%%%%%%%%%%%%%%%%%%%%%%%%%%%%%
The motivation for developing the perturbation theory is to isolate and capture the effect of the host absorption or host gain on the overall extinction, which is not possible within the MLWA (the latter captures only overall extinction).
To this end, a useful parametrization of the relative dielectric function $\veps$ between the sphere and the host, suitable for studying the departure from nonabsorbing host, is (cf. Eq.~(9) of ref.~\citenum{Zhang2022b}):
\bg
\veps =\fr{\veps_1}{\veps_2}=\fr{(n_1'+in_1'')^2}{(n_2'+in_2'')^2}
=\fr{(n_1'/n_2'+in_1''/n_2')^2}{(1+in_2''/n_2')^2}
=
\fr{\veps_t}{(1+i\bar\kappa)^2},
\lb{kappa}
\eg
where $\veps_t=(n_1'/n_2'+in_1''/n_2')^2=\veps_1/(n_2')^2$ is the relative (in general complex number if $\veps_1$ is complex) dielectric function between the sphere and a {\em nonabsorbing} host with real refractive index $n_2'$, and $\bar\kappa=n_2''/n_2'$ is a normalized extinction coefficient representing the magnitude of the host dissipation. 
On substituting the Taylor expansion of the electric dipole term $a_1$,
\bg
a_1(\veps)=a_1(\veps_t) + \bar\kappa \left. \fr{{\rm d}a_1}{{\rm d}\bar\kappa}\right|_{\veps=\veps_t} + {\cal O}(\bar\kappa^2),
\nonumber 
\eg
in the expression Eq.~\eqref{Cext} of the apparent cross section, $C_{\rm ext}$, it is in principle possible to provide in a systematic way the results for the extinction efficiency, $Q_{\rm ext}=C_{\rm ext}/\pi a^2$, in the first order of $\bar\kappa$ for both gain and absorbing media in the dipole approximation.
The derivative ${\rm d}a_1/{\rm d}\bar\kappa$ is determined by Eq.~S21 from Section~S5 in Supplement 1,
\bea
\fr{{\rm d}a_1}{{\rm d}\bar\kappa}  =  - 12n_2' x' \bar\kappa\, 
 \mb{Re}\left[\fr{ 2- \veps_t(\veps_t-1) + \fr{1}{5}(\veps_t^2-\veps_t +2) (x')^2}{n_2 D^2(\veps_t,x')}\right], 
\lb{a1dm}
\eea
where $x'=x_0n_2'$, with $x_0=2\pi a /\lambda$ being the size parameter in vacuum host, and
\bg
D(\veps_t,x') = \veps_t +2 -3(\veps_t-2)(x')^2/5 -2i (\veps_t-1)(x')^3/3
\lb{si4}
\eg
is the denominator $D$ in Eq.~(\rf{mlwa1}) in the limit $\bar\kappa\to 0$. One finds eventually
\bea
Q_{\rm ext} &=& \fr{C_{\rm ext}}{\pi a^2} \approx \fr{2}{x'}\, \mb{Re}\left\{  \fr{3}{x}\, \left[ a_1(\veps_t) + \bar\kappa \left. \fr{da_1}{d\bar\kappa}\right|_{\veps=\veps_t} \right]
\right\}
\nn\\
&=& \fr{2}{x'} \mb{Re} \left\{\fr{3}{x} a_1 (\veps_{t}) \right\} -
\nn\\
&&
12n_2' x' \bar\kappa\, \mb{Re}\left[\fr{ 2- \veps_t(\veps_t-1) + \fr{1}{5}(\veps_t^2-\veps_t +2) (x')^2}{n_2 D^2(\veps_t,x')} \right],\hst{1cm}
\lb{Cextl1}
\eea
which defines the first order expansion of $Q_{\rm ext}$ in the parameter $\bar\kappa$ within the perturbation theory (PT). The first term on the rhs is the dipole MLWA in a nonabsorbing host characterized by $\veps_{t}$. The second term  on the rhs is the perturbation correction.
In what follows, we will refer to Eq.~\eqref{Cextl1} as the $\bar\kappa$-PT approximation.

\section{Results}
\lb{sc:res}
%%%%%%%%%%%%%%%%%%%%%%%%%%%%%

\subsection{The dipole MLWA vs perturbation theory }
\lb{sc:resmlwa}
%%%%%%%%%%%%%%%%%%%%%%%%%%%%%
\begin{figure*}[!ht]
\centering
\includegraphics[width=\linewidth]{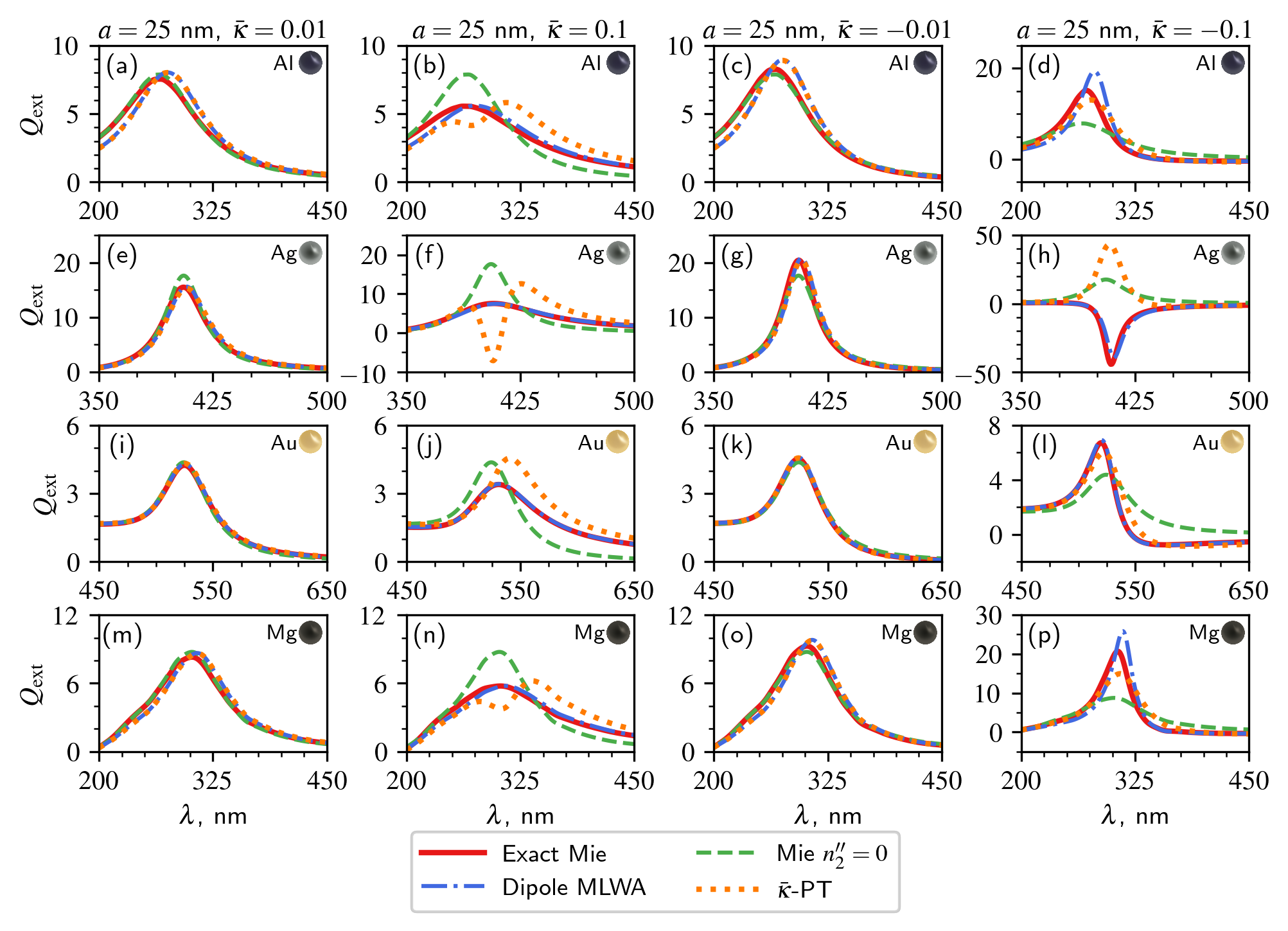}
\caption{Extinction spectra, $Q_{\rm ext}$, for Al, Ag, Au and Mg spherical nanoparticles with $a=25$~nm embedded in a host absorption medium with $n_2'=1.33$ (left two columns) and gain host medium (right two columns) for $\bar\kappa=0.01, 0.1$ ($\bar\kappa=-0.01, -0.1$).
$Q_{\rm ext}$ are shown as calculated by Eq.~(\ref{Cext}) of the Mie theory (solid red line), within dipole MLWA of Eq.~(\ref{mlwa1}) (blue dot-dashed line), the Mie theory approximation with the nonabsorbing host (green dashed line), and by the $\bar\kappa$-PT given by Eq.~(\ref{Cextl1}) (dotted orange line).
The refractive indices of Ag, Al, Au were taken from ref.~\citenum{McPeak2015} and those of Mg from ref.~\citenum{Palm2018}.
Whereas absorption decreases the height of a LSPR and broadens its width, the gain does just the opposite.
Note enlarged scale on the vertical axis in the last column for $\bar\kappa=-0.1$.}
\label{fig:mlwa_ext_abs_gain}
\end{figure*}
%%%%%%%%%%%%%%%%%%%%%%%%%

Fig.~\ref{fig:mlwa_tau} compares performance of various approximations relative to the exact Mie theory involving
\begin{itemize}

    \item the dipole MLWA (with the sole term $T_{E1}$ in Eq.~(\rf{Cext}) given by Eq.~(\rf{mlwa1}))
    
    \item the Mie theory approximation with nonabsorbing host (determined by Eq.~(\rf{Cext}) with $n_2''=0$)
    
    \item the first-order  $\bar\kappa$-PT given by Eq.~(\ref{Cextl1})

\end{itemize}
%%%%%%%%%%%%%
in water-like host ($n_2'=1.33$) for different host absorption characterized by different values of $\bar\kappa$ on the example of Ag spheres with radii between $10$ and $50$~nm. 
Fig.~\ref{fig:mlwa_ext_abs_gain} compares results for spheres made from most common plasmonic materials Al, Ag, Au, Mg with radius $a=25$~nm in water-like host ($n_2'=1.33$) with $\bar\kappa=\pm0.01$ and $\pm0.1$ (similar results for glass-like host ($n_2'=1.5$) are presented in Fig.~S2 in Supplement 1).
For different metals of plasmonic particles tabulated data of refractive indices have been used as follows: Al~\cite{McPeak2015}, Ag~\cite{McPeak2015}, Au~\cite{McPeak2015}, Mg~\cite{Palm2018}.
As demonstrated in those figures, the dipole MLWA can be very precise.
In the case of Ag and Au nanoparticles, the dipole MLWA essentially overlies with the exact Mie theory results for $a\lesssim 25$~nm.
The agreement, slightly better in the case of Au than Ag, continues to be acceptable up to $a\sim 50$ nm (Fig.~S1III in Supplement 1), and can be used, at least qualitatively, up to $a\sim 70$~nm (Fig.~S1IV in Supplement 1).
In the case of Al and Mg nanoparticles, a slight deviation from Mie theory results becomes visible by the naked eye already for $a\gtrsim 25$~nm (cf. Figs.~S1I and S1II in Supplement 1).
The agreement continues to be acceptable up to $a\sim 40$~nm and can be used at least qualitatively up to $a\sim 50$ nm (Fig.~S1III in Supplement 1).

A first general observation is that with increasing sphere radius the performance of the dipole MLWA worsens. This is not surprising, as the MLWA is by definition is a long wavelength approximation. The same applies to the $\bar\kappa$-PT. 
The $\bar\kappa$-PT overlies with the dipole MLWA for all $a\le 50$~nm, provided that $\bar\kappa\le 0.01$.
The latter justifies a posteriori that our $\bar\kappa$-PT is correct.
Another interesting tendency revealed by Fig.~\ref{fig:mlwa_tau} (Figs.~S1I and S1II in Supplement 1) is that the maximal radius $a$ for which the dipole MLWA remains accurate increases with increasing $\bar\kappa$.
This observation can be explained by the fact that for a given particle radius an increasing host absorption
suppresses the contribution of higher-order multipoles that would
normally arise at larger particle sizes. This allows the dipole MLWA to remain valid over a wider range of particle sizes, as the multipole effects are suppressed in absorbing media.

%%%%%%%%%%%%%%%%%%%%%%%%
\begin{figure}[!hb]
\centering
\includegraphics[width=3in]{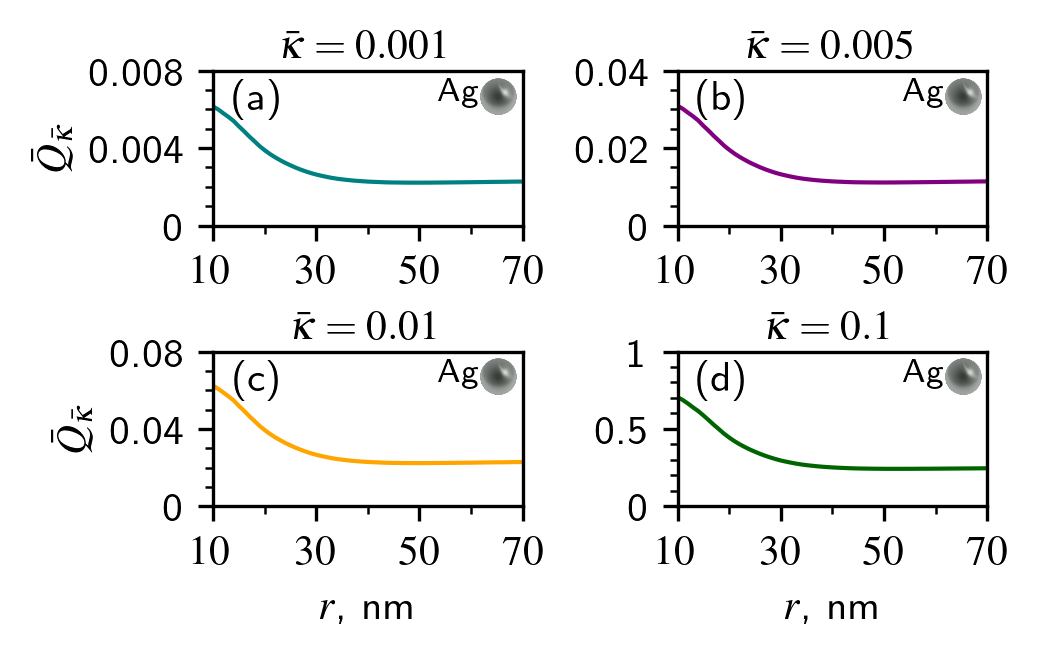}
\caption{Size dependence of $\bar Q_{\bar \kappa}$, the ratio of the PT correction to the leading term on the rhs of Eq.~(\ref{Cextl1}), for Ag sphere at the (size-dependent) LSPR wavelength for different $\bar\kappa$: (a) $0.001$, (b) $0.005$, (c) $0.01$ and (d) $0.1$.}
\label{fig:mlwa_sz}
\end{figure}
%%%%%%%%%%%%%%%%%%%%%%%% 

%%%%%%%%%%%%%%%%%%
\begin{figure}
\centering
\includegraphics[width=3in]{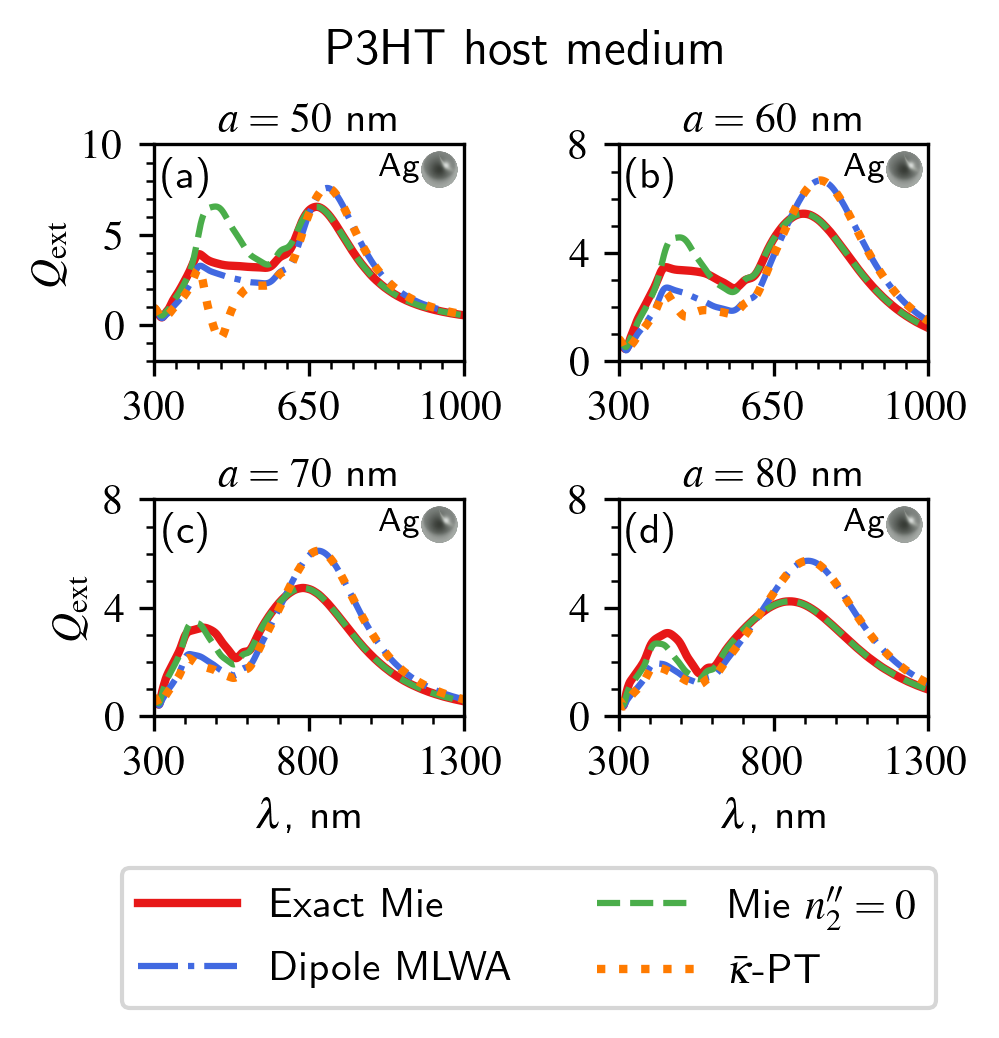} 
\caption{
Extinction spectra, $Q_{\rm ext}$, for spherical Ag nanoparticles with (a) $a=50$~nm, (b) $a=60$~nm, (c) $a=70$~nm, and (d) $a=80$~nm embedded in Poly(3-hexylthiophene) (P3HT) host medium.
Spectra are shown as calculated with the exact Mie theory (solid red line), the Mie theory for non-absorbing media ($n_2''=0$) (green dashed line), the dipole MLWA (blue dot-dashed line), and the $\bar\kappa$-PT (dotted orange line).}
\label{fig:mlwa_p3ht}
\end{figure}
%%%%%%%%%%%%%%%%%%
Importantly, the results of $\bar\kappa$-PT and the dipole MLWA overlie with the exact Mie theory results for $a\lesssim 25$~nm, which supports earlier observation on the range of validity of the MLWA in an absorbing host by Khlebtsov~\cite{Khlebtsov2021}.
Whereas the dipole MLWA continues to overlie with the exact Mie theory results for $a\lesssim 25$ nm irrespective of $\bar\kappa$, the results of the $\bar\kappa$-PT begin to deviate from those of the dipole MLWA for $\bar\kappa\gtrsim 0.05$. 
An extreme case of such a deviation is provided for Ag sphere with $a=25$ nm and $\bar\kappa= \pm 0.1$ shown in Fig.~\ref{fig:mlwa_ext_abs_gain}(f),(h). Whereas the dipole MLWA nearly overlies the exact result showing a positive (negative) extinction for $\bar\kappa= 0.1$ ($\bar\kappa=- 0.1$), the $\bar\kappa$-PT in the respective cases indicates a negative (positive) extinction.
This is to be expected, because the $\bar\kappa$-PT is a first order perturbation theory of the dipole MLWA in $\bar\kappa$ and is expected to eventually break down above a certain threshold value of 
$|\bar\kappa|$. Negative extinction shown in Fig.~\ref{fig:mlwa_ext_abs_gain}(h) is an indication of that particle losses have been more than compensated by the gain medium - a precursor of lasing action~\cite{Bergman2003,Lawandy2004,Veltri2016}.

Surprisingly, the smaller the particle, the greater the deviation of the $\bar\kappa$-PT relative to the dipole MLWA for $\bar\kappa\gtrsim 0.05$ (see Fig.~S1 and S2 in Supplement 1). 
The origin of this behaviour is that $D(\veps_t)$ is typically small in a proximity of a LSPR (cf. the generalized Fröhlich condition Eq.~(\rf{vepsz})).
Whereas the first term in Eq.~(\rf{Cextl1}) is of the order $1/D(\veps_t)$, the term proportional to $\bar\kappa$ is of the order $1/D^2(\veps_t)$. However, when the size parameter $x$ increases, $|D(\veps_t)|$ at a LSPR decreases (due to a larger separation from the complex zero), and, as illustrated in Fig. \ref{fig:mlwa_sz}, $\bar Q_{\bar \kappa}$, defined as the ratio of the PT correction to the leading term on the rhs of Eq.~(\ref{Cextl1}), gradually decreases, 
whereby the first-order $\bar\kappa$-PT begins to approximate the MLWA. 
Note also how the ratio $\bar Q_{\bar \kappa}$ increases with $\bar\kappa$, which, as expected, explains why the $\bar\kappa$-PT
begins to deviate from the MLWA with increasing $\bar\kappa$.

\subsection{Mie theory approximation with nonabsorbing host}
\lb{sc:resmie}
%%%%%%%%%%%%%%%%%%%%%%%%%%%%%%%%%%%%%%%%%%%%
Similar behaviour is observed for the Mie theory approximation with nonabsorbing host ($n_2''=0$) that begins to deviate from the exact Mie theory results, but somewhat earlier, beginning with $\bar\kappa= 0.01$.
Again the smaller the particle, the greater the deviation.
Nevertheless, in a relatively weakly absorbing host the Mie theory approximation with nonabsorbing host becomes the best approximation. For example, the Mie theory in nonabsorbing host begins to overlie with the Mie theory in an absorbing host for $a\gtrsim 50$ nm and $\bar\kappa = 0.01$ (see Fig.~S1 in Supplement 1).
A threshold radius for which it happens increases with $\bar\kappa$. 

Why the Mie theory approximation with nonabsorbing host becomes the best approximation for sufficiently large $a$ can be explained by increasing relevance of higher-order multipole contributions with increasing particle size.
This is understandable, because with increasing $a$ higher order multipoles, which are obviously absent in any dipole approximation, become more and more relevant. 
This is demonstrated also in Fig.~\ref{fig:mlwa_p3ht} showing the effect of increasing the radius of spherical Ag nanoparticles in Poly(3-hexylthiophene) (P3HT) host medium, which was used in a number of recent studies~\cite{Peck2019,Peck2021,Khlebtsov2021,Zhang2022b}.
Unlike inherent dipole approximations, the Mie theory approximation with nonabsorbing host captures reasonably well both the dipole and quadrupole peaks. 
Obviously, higher-order MLWA of ref.~\citenum{Rasskazov2021} could have captured the quadrupole peak, but this goes beyond the scope of present study. 

%%% The fact that the model used in the manuscript can be safely used up to the laser threshold has been verified both experimentally and theoretically more than 15 years ago in Ref.~\cite{vanderMolen2006}.

\subsection{Isolating the host dissipative effects on the extinction cross sections}
\begin{figure}[!ht]
\centering
\includegraphics[width=3.1in]{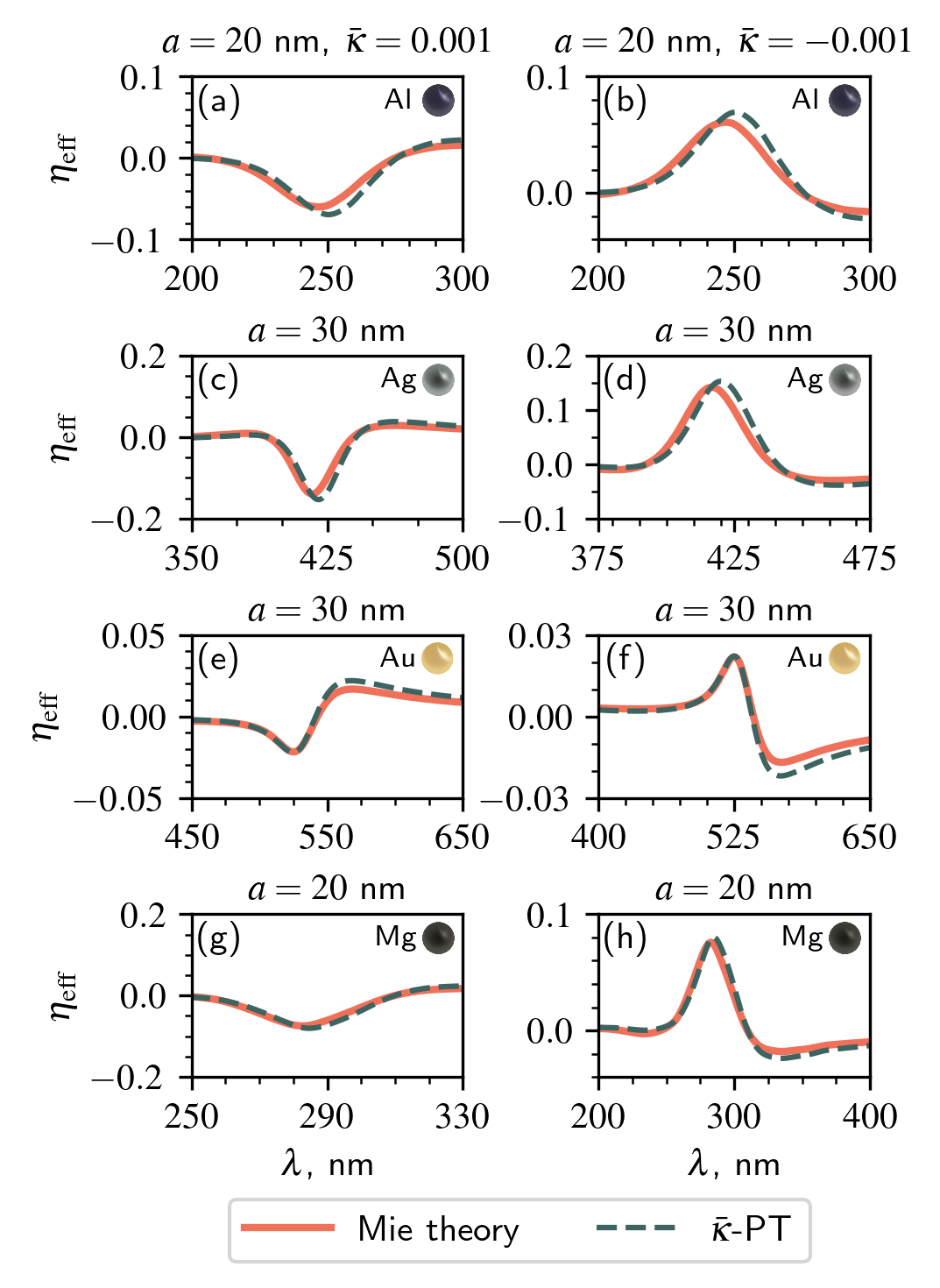}
\caption{The effect $\eta_{\rm eff}$ of the host absorption (left column) and the gain host medium (right column) on $Q_{\rm ext}$ in the Mie theory (solid red line) and the $\bar\kappa$-PT (dashed dark green line) for Al, Ag, Au and Mg materials with different radii $a$ embedded in the host medium with
$n_2'=1.33$ and $\bar\kappa=0.001$ (left column) and
$\bar\kappa=-0.001$ (right column).}
\label{fig:mlwa_ef}
\end{figure}

%%%%%%%%%%%%%%%%%%%%%%%%%%%%
The dissipative effects of the host on the extinction cross sections, $Q_{\rm ext}$, can obviously be isolated by subtracting from the exact value $Q_{\rm ext}=Q_{{\rm ext};\veps}$ the value of $Q_{{\rm ext};\veps_t}$ obtained by the Mie theory for the non-absorbing host characterised by $n_2''=0$ (i.e. $\veps\to \veps_t$).
In what follows, we denote the difference by $\eta_{\rm eff}:=Q_{{\rm ext};\veps} - Q_{{\rm ext};\veps_t}$.
Thus $\eta_{\rm eff}$ quantifies the contribution of the host absorption or gain in the resulting $Q_{\rm ext}$.
To our satisfaction, it turns out that, in a suitable parameter range, the analytic $\bar\kappa$-PT can reliably capture the effect of the host absorption on the extinction efficiency of a plasmonic nanosphere as demonstrated in Fig.~\ref{fig:mlwa_ef}.
Not surprisingly, the agreement can also be reached in the case of gain media, which opens door for analysis of promising applications involving active media, see Fig.~\ref{fig:mlwa_ext_abs_gain} (Fig.~S5 in Supplement 1) such as spasers~\cite{Bergman2003,Lawandy2004,Veltri2016}.
Therefore, within the range of its validity, the first order $\bar\kappa$-PT allows one to both intuitively and analytically understand the mechanisms of host dissipation or gain on the extinction efficiency of a plasmonic nanosphere. 
%%%%%%%%%%%%%%%%%%%
%%%%%%%%%%%%%%%%%%%%%%%%%%%%%

\section{Discussion}
\lb{sc:disc}
%%%%%%%%%%%%%%%%%%%%%%%%%%%%%%%%%%%%%%%%%%%%%%%%%%
First-principles far-field computations based on the general Lorenz–Mie theory showed that increasing absorption in the host medium {\em broadens} and suppresses plasmon resonances in the apparent extinction~\cite{Mishchenko2019}. 
Such a broadening and suppression of plasmon resonances is also present in all approximation shown in Figs.~\ref{fig:mlwa_tau} and ~\ref{fig:mlwa_ext_abs_gain}. The effect of absorption in the host thus goes in the same direction as increasing absorption within the particle.
Noteworthy, increasing gain in the host medium {\em narrows} plasmon resonances and increases their amplitude in the apparent extinction as demonstrated in Fig.~\ref{fig:mlwa_ext_abs_gain} (cf. also Fig.~S3). This goes along physical intuition.
Contrary to the apparent extinction, in the case of the inherent cross-sections a surrounding lossy medium was shown to {\em narrow} the plasmonic resonances, as well as {\em increase} their amplitude dramatically~\cite{Peck2019,Khlebtsov2021}.

The dipole MLWA can, in principle, be used up to the laser threshold associated with the so-called spectral singularity of non-Hermitian models. Indeed, as it has been verified both experimentally and theoretically more than 15 years ago, a description of gain media consisting in changing the sign of extinction is a valid approximation up to the laser threshold~\cite{vanderMolen2006}. This could be particularly important for applications in spasers and nano-lasers.

For the sake of completeness, rather exhaustive MLWA analysis in the absorbing case has been recently provided by Khlebtsov~\cite{Khlebtsov2021} by employing slightly different MLWA. 
The study of ref.~\citenum{Khlebtsov2021} arrived at similar conclusions that MLWA is very precise for $a\lesssim 25$~nm.
Our study complements ref.~\citenum{Khlebtsov2021} with 
(i) a first order analytic perturbation theory around a nonabsorbing host in a normalized extinction coefficient $\bar\kappa$ ($\bar\kappa$-PT) and 
(ii) an investigation of gain media.
In addition, we have examined the behaviour of Al and Mg nanoparticles.

\section{Conclusions}
\lb{sc:conc}
%%%%%%%%%%%%%%%%%%%%%%
Our previously developed modified dipole long-wave approximation (MLWA) was shown to essentially overly with the exact Mie theory results for $a\lesssim 25$~nm ($a\lesssim 20$~nm) in the case of Ag and Au (Al and Mg) nanoparticles.
The agreement for Au and Ag (Al and Mg) nanoparticles, slightly better in the case of Au than Ag, continues to be acceptable up to $a\sim 50$~nm ($a\sim 40$~nm), and can be used, at least qualitatively, up to $a\sim 70$~nm ($a\sim 50$~nm).
Expanding the discussion to the case of larger nanoparticles with $a>70$~nm almost certainly requires consideration of MLWA for higher order multipoles~\cite{Schebarchov2013,Rasskazov2021}.
We developed within the dipole MLWA a first order analytic perturbation theory (PT) around a nonabsorbing host in a normalized extinction coefficient $\bar\kappa$ and investigated its properties.
It was shown that, in a suitable parameter range, the $\bar\kappa$-PT can reliably isolate and capture the effect of host absorption or host gain on the overall extinction efficiency of spherical plasmonic nanoparticles. 
Considering growing interest in light-matter interactions, we expect that our results will help in designing optimal systems comprising plasmonic nanoparticles embedded in suitable dissipative or gain media for various applications, such as photothermal therapy~\cite{Jain2012, Dreaden2012}, spasers and other active medium devices~\cite{Bergman2003, Lawandy2004, Veltri2016}.

\section{Funding}
% Content in the funding section will be generated entirely from details submitted to Prism. Authors may add placeholder text in the manuscript to assess length, but any text added to this section in the manuscript will be replaced during production and will display official funder names along with any grant numbers provided. If additional details about a funder are required, they may be added to the Acknowledgments, even if this duplicates information in the funding section. See the example below in Acknowledgements. For preprint submissions, please include funder names and grant numbers in the manuscript.

The analytical study of absorbing media is funded by Russian Science Foundation grant № 23-19-00511.

\section{Declaration of competing interest}
There are no conflicts to declare.

\section{Data availability} Data underlying the results presented in this paper are not publicly available at this time but may be obtained from the authors upon reasonable request.

\section{Supplemental document}
See Supplement materials for supporting content. 

%%%%%%%%%% If using BibTeX:
\bibliographystyle{unsrt} 
\bibliography{MLWA_abs_host}

\begin{thebibliography}{10}

\bibitem{Newton1982}
Roger~G. Newton.
\newblock {\em Scattering Theory of Waves and Particles}.
\newblock Springer Berlin Heidelberg, Berlin, Heidelberg, 1982.

\bibitem{Bohren1998}
Craig~F. Bohren and Donald~R. Huffman.
\newblock {\em Absorption and Scattering of Light by Small Particles}.
\newblock Wiley-VCH Verlag GmbH \& Co. KGaA, April 1998.

\bibitem{Fuchs1968}
Ronald Fuchs and K.~L. Kliewer.
\newblock Optical modes of vibration in an ionic crystal sphere.
\newblock {\em J. Opt. Soc. Am.}, 58(3):319--330, March 1968.

\bibitem{Mundy1974}
W.~C. Mundy, J.~A. Roux, and A.~M. Smith.
\newblock Mie scattering by spheres in an absorbing medium*.
\newblock {\em J. Opt. Soc. Am.}, 64(12):1593--1597, December 1974.

\bibitem{Chylek1979}
Petr Ch{\'y}lek and R.~G. Pinnick.
\newblock Nonunitarity of the light scattering approximations.
\newblock {\em Appl. Opt.}, 18(8):1123--1124, April 1979.

\bibitem{Bohren1979}
Craig~F Bohren and Daya~P Gilra.
\newblock Extinction by a spherical particle in an absorbing medium.
\newblock {\em J. Colloid Interface Sci.}, 72(2):215--221, November 1979.

\bibitem{Lebedev1999}
A~N Lebedev, M~Gartz, U~Kreibig, and O~Stenzel.
\newblock Optical extinction by spherical particles in an absorbing medium: Application to composite absorbing films.
\newblock {\em Eur. Phys. J. D.}, 6(3):365--373, 1999.

\bibitem{Sudiarta2001}
I.~Wayan Sudiarta and Petr Chylek.
\newblock Mie-scattering formalism for spherical particles embedded in an absorbing medium.
\newblock {\em J. Opt. Soc. Am. A}, 18(6):1275--1278, June 2001.

\bibitem{Yang2002}
Ping Yang, Bo-Cai Gao, Warren~J. Wiscombe, Michael~I. Mishchenko, Steven~E. Platnick, Hung-Lung Huang, Bryan~A. Baum, Yong~X. Hu, Dave~M. Winker, Si-Chee Tsay, and Seon~K. Park.
\newblock Inherent and apparent scattering properties of coated or uncoated spheres embedded in an absorbing host medium.
\newblock {\em Appl. Opt.}, 41(15):2740--2759, May 2002.

\bibitem{Fu2006}
Qiang Fu and Wenbo Sun.
\newblock Apparent optical properties of spherical particles in absorbing medium.
\newblock {\em J. Quant. Spectrosc. Radiat. Transf.}, 100(1-3):137--142, July 2006.

\bibitem{Mishchenko2007}
Michael~I. Mishchenko.
\newblock Electromagnetic scattering by a fixed finite object embedded in an absorbing medium.
\newblock {\em Opt. Express}, 15(20):13188--13201, 2007.

\bibitem{Mishchenko2017}
Michael~I. Mishchenko, Gorden Videen, and Ping Yang.
\newblock Extinction by a homogeneous spherical particle in an absorbing medium.
\newblock {\em Opt. Lett.}, 42(23):4873--4876, December 2017.

\bibitem{Mishchenko2018}
Michael~I. Mishchenko and Ping Yang.
\newblock Far-field lorenz{\textendash}mie scattering in an absorbing host medium: Theoretical formalism and fortran program.
\newblock {\em J. Quant. Spectrosc. Radiat. Transf.}, 205:241--252, January 2018.

\bibitem{Mishchenko2019}
Michael~I. Mishchenko and Janna~M. Dlugach.
\newblock Multiple scattering of polarized light by particles in an absorbing medium.
\newblock {\em Appl. Opt.}, 58(18):4871--4877, June 2019.

\bibitem{Peck2019}
Ryan~L. Peck, Alexandre~G. Brolo, and Reuven Gordon.
\newblock Absorption leads to narrower plasmonic resonances.
\newblock {\em J. Opt. Soc. Am. B}, 36(8):F117--F122, August 2019.

\bibitem{Mishchenko2019b}
Michael~I. Mishchenko, Maxim~A. Yurkin, and Brian Cairns.
\newblock Scattering of a damped inhomogeneous plane wave by a particle in a weakly absorbing medium.
\newblock {\em OSA Continuum}, 2(8):2362--2368, August 2019.

\bibitem{Khlebtsov2021}
Nikolai~G. Khlebtsov.
\newblock Extinction, absorption, and scattering of light by plasmonic spheres embedded in an absorbing host medium.
\newblock {\em Phys. Chem. Chem. Phys.}, 23(40):23141--23157, 2021.

\bibitem{Dong2021}
Jian Dong, Wenjie Zhang, and Linhua Liu.
\newblock Discrete dipole approximation method for electromagnetic scattering by particles in an absorbing host medium.
\newblock {\em Opt. Express}, 29(5):7690--7705, March 2021.

\bibitem{Zhang2022b}
Shangyu Zhang, Jian Dong, Wenjie Zhang, {Minggang Luo}, and Linhua Liu.
\newblock Extinction by plasmonic nanoparticles in dispersive and dissipative media.
\newblock {\em Opt. Lett.}, 47(21):5577--5580, November 2022.

\bibitem{Mishchenko2019a}
Michael~I. Mishchenko and Janna~M. Dlugach.
\newblock Plasmon resonances of metal nanoparticles in an absorbing medium.
\newblock {\em OSA Continuum}, 2(12):3415--3421, December 2019.

\bibitem{Videen2003}
Gorden Videen and Wenbo Sun.
\newblock Yet another look at light scattering from particles in absorbing media.
\newblock {\em Appl. Opt.}, 42(33):6724--6727, 2003.

\bibitem{Meier1983}
M.~Meier and A.~Wokaun.
\newblock Enhanced fields on large metal particles: dynamic depolarization.
\newblock {\em Opt. Lett.}, 8(11):581--583, November 1983.

\bibitem{Zeman1984}
Ellen~J. Zeman and George~C. Schatz.
\newblock Electromagnetic theory calculations for spheroids: An accurate study of the particle size dependence of sers and hyper-raman enhancements.
\newblock In Bernard Pullman, Joshua Jortner, Abraham Nitzan, and Benjamin Gerber, editors, {\em Dynamics on Surfaces}, volume~17, pages 413--424, Dordrecht, 1984. Springer Netherlands.

\bibitem{Zeman1987}
Ellen~J. Zeman and George~C. Schatz.
\newblock An accurate electromagnetic theory study of surface enhancement factors for silver, gold, copper, lithium, sodium, aluminum, gallium, indium, zinc, and cadmium.
\newblock {\em J. Phys. Chem.}, 91(3):634--643, January 1987.

\bibitem{Kelly2003}
K.~Lance Kelly, Eduardo Coronado, Lin~Lin Zhao, and George~C. Schatz.
\newblock The optical properties of metal nanoparticles: the influence of size, shape, and dielectric environment.
\newblock {\em J. Phys. Chem. B}, 107(3):668--677, January 2003.

\bibitem{Kuwata2003}
Hitoshi Kuwata, Hiroharu Tamaru, Kunio Esumi, and Kenjiro Miyano.
\newblock Resonant light scattering from metal nanoparticles: Practical analysis beyond rayleigh approximation.
\newblock {\em Appl. Phys. Lett.}, 83(22):4625--4627, December 2003.

\bibitem{Moroz2009}
Alexander Moroz.
\newblock Depolarization field of spheroidal particles.
\newblock {\em J. Opt. Soc. Am. B}, 26(3):517--527, March 2009.

\bibitem{Zoric2011}
Igor Zori{\'c}, Michael Z{\"a}ch, Bengt Kasemo, and Christoph Langhammer.
\newblock Gold, platinum, and aluminum nanodisk plasmons: Material independence, subradiance, and damping mechanisms.
\newblock {\em ACS Nano}, 5(4):2535--2546, April 2011.

\bibitem{Massa2013}
Enrico Massa, Stefan~A Maier, and Vincenzo Giannini.
\newblock An analytical approach to light scattering from small cubic and rectangular cuboidal nanoantennas.
\newblock {\em New J. Phys.}, 15(6):063013, June 2013.

\bibitem{LeRu2013}
Eric~C. Le~Ru, Walter R.~C. Somerville, and Baptiste Augui{\'e}.
\newblock Radiative correction in approximate treatments of electromagnetic scattering by point and body scatterers.
\newblock {\em Phys. Rev. A}, 87(1):012504, January 2013.

\bibitem{Schebarchov2013}
Dmitri Schebarchov, Baptiste Augui{\'e}, and Eric~C. Le~Ru.
\newblock Simple accurate approximations for the optical properties of metallic nanospheres and nanoshells.
\newblock {\em Phys. Chem. Chem. Phys.}, 15(12):4233--4242, 2013.

\bibitem{Januar2020}
Mochamad Januar, Bei Liu, Jui-Ching Cheng, Koji Hatanaka, Hiroaki Misawa, Hui-Hsin Hsiao, and Kou-Chen Liu.
\newblock Role of depolarization factors in the evolution of a dipolar plasmonic spectral line in the far- and near-field regimes.
\newblock {\em J. Phys. Chem. C}, 124(5):3250--3259, February 2020.

\bibitem{Rasskazov2020a}
Ilia~L. Rasskazov, P.~Scott Carney, and Alexander Moroz.
\newblock Intriguing branching of the maximum position of the absorption cross section in mie theory explained.
\newblock {\em Opt. Lett.}, 45(14):4056--4059, July 2020.

\bibitem{Rasskazov2021}
Ilia~L. Rasskazov, Vadim~I. Zakomirnyi, Anton~D. Utyushev, P.~Scott Carney, and Alexander Moroz.
\newblock Remarkable predictive power of the modified long wavelength approximation.
\newblock {\em J. Phys. Chem. C}, 125(3):1963--1971, January 2021.

\bibitem{McPeak2015}
Kevin~M. {McPeak}, Sriharsha~V. Jayanti, Stephan J.~P. Kress, Stefan Meyer, Stelio Iotti, Aurelio Rossinelli, and David~J. Norris.
\newblock Plasmonic films can easily be better: Rules and recipes.
\newblock {\em {ACS} Photonics}, 2(3):326--333, 2015.

\bibitem{Palm2018}
Kevin~J. Palm, Joseph~B. Murray, Tarun~C. Narayan, and Jeremy~N. Munday.
\newblock Dynamic optical properties of metal hydrides.
\newblock {\em {ACS} Photonics}, 5(11):4677--4686, 2018.

\bibitem{Bergman2003}
David~J. Bergman and Mark~I. Stockman.
\newblock Surface plasmon amplification by stimulated emission of radiation: Quantum generation of coherent surface plasmons in nanosystems.
\newblock {\em Phys. Rev. Lett.}, 90(2):027402, January 2003.

\bibitem{Lawandy2004}
N.~M. Lawandy.
\newblock Localized surface plasmon singularities in amplifying media.
\newblock {\em Appl. Phys. Lett.}, 85(21):5040--5042, November 2004.

\bibitem{Veltri2016}
Alessandro Veltri, Arkadi Chipouline, and Ashod Aradian.
\newblock Multipolar, time-dynamical model for the loss compensation and lasing of a spherical plasmonic nanoparticle spaser immersed in an active gain medium.
\newblock {\em Sci. Rep.}, 6(1):33018, September 2016.

\bibitem{Peck2021}
Ryan Peck, Ali Khademi, Juanjuan Ren, Stephen Hughes, Alexandre~G. Brolo, and Reuven Gordon.
\newblock Plasmonic linewidth narrowing by encapsulation in a dispersive absorbing material.
\newblock {\em Phys. Rev. Research}, 3(1):013014, January 2021.

\bibitem{vanderMolen2006}
Karen~L. van~der Molen, Peter Zijlstra, Ad~Lagendijk, and Allard~P. Mosk.
\newblock Laser threshold of mie resonances.
\newblock {\em Opt. Lett.}, 31(10):1432--1434, May 2006.

\bibitem{Jain2012}
Prashant~K Jain, Ivan~H El-Sayed, and Mostafa~A El-Sayed.
\newblock Plasmonic photothermal therapy (pptt) using gold nanoparticles.
\newblock {\em Lasers in Medical Science}, 23(3):217--228, 2012.

\bibitem{Dreaden2012}
Erik~C Dreaden, Alaaldin~M Alkilany, Xiaohua Huang, Catherine~J Murphy, and Mostafa~A El-Sayed.
\newblock The golden age: gold nanoparticles for biomedicine.
\newblock {\em Chemical Society Reviews}, 41(7):2740--2779, 2012.

\end{thebibliography}


\begin{thebibliography}{1}

\bibitem{SI_Zhang2022b}
Shangyu Zhang, Jian Dong, Wenjie Zhang, {Minggang Luo}, and Linhua Liu.
\newblock Extinction by plasmonic nanoparticles in dispersive and dissipative media.
\newblock {\em Opt. Lett.}, 47(21):5577--5580, November 2022.

\bibitem{SI_Bohren1998}
Craig~F. Bohren and Donald~R. Huffman.
\newblock {\em Absorption and Scattering of Light by Small Particles}.
\newblock Wiley-VCH Verlag GmbH \& Co. KGaA, April 1998.

\bibitem{SI_Rasskazov2020a}
Ilia~L. Rasskazov, P.~Scott Carney, and Alexander Moroz.
\newblock Intriguing branching of the maximum position of the absorption cross section in mie theory explained.
\newblock {\em Opt. Lett.}, 45(14):4056--4059, July 2020.

\bibitem{SI_Rasskazov2021}
Ilia~L. Rasskazov, Vadim~I. Zakomirnyi, Anton~D. Utyushev, P.~Scott Carney, and Alexander Moroz.
\newblock Remarkable predictive power of the modified long wavelength approximation.
\newblock {\em J. Phys. Chem. C}, 125(3):1963--1971, January 2021.

\bibitem{SI_Newton1982}
Roger~G. Newton.
\newblock {\em Scattering Theory of Waves and Particles}.
\newblock Springer Berlin Heidelberg, Berlin, Heidelberg, 1982.

\bibitem{SI_Bohren1979}
Craig~F Bohren and Daya~P Gilra.
\newblock Extinction by a spherical particle in an absorbing medium.
\newblock {\em J. Colloid Interface Sci.}, 72(2):215--221, November 1979.

\bibitem{SI_Mishchenko2018}
Michael~I. Mishchenko and Ping Yang.
\newblock Far-field lorenz{\textendash}mie scattering in an absorbing host medium: Theoretical formalism and fortran program.
\newblock {\em J. Quant. Spectrosc. Radiat. Transf.}, 205:241--252, January 2018.

\end{thebibliography}

\end{document}

\maketitle
\newpage

\section{Supplementary figures}
\lb{sec:sfgrs}
%%%%%%%%%%%%%%%%%%%%%%%%%%%%%%%%%
\begin{figure}[!htbp]
\centering
    \begin{subfigure}{0.358\textwidth}
        \centering
        \includegraphics[width=\linewidth]{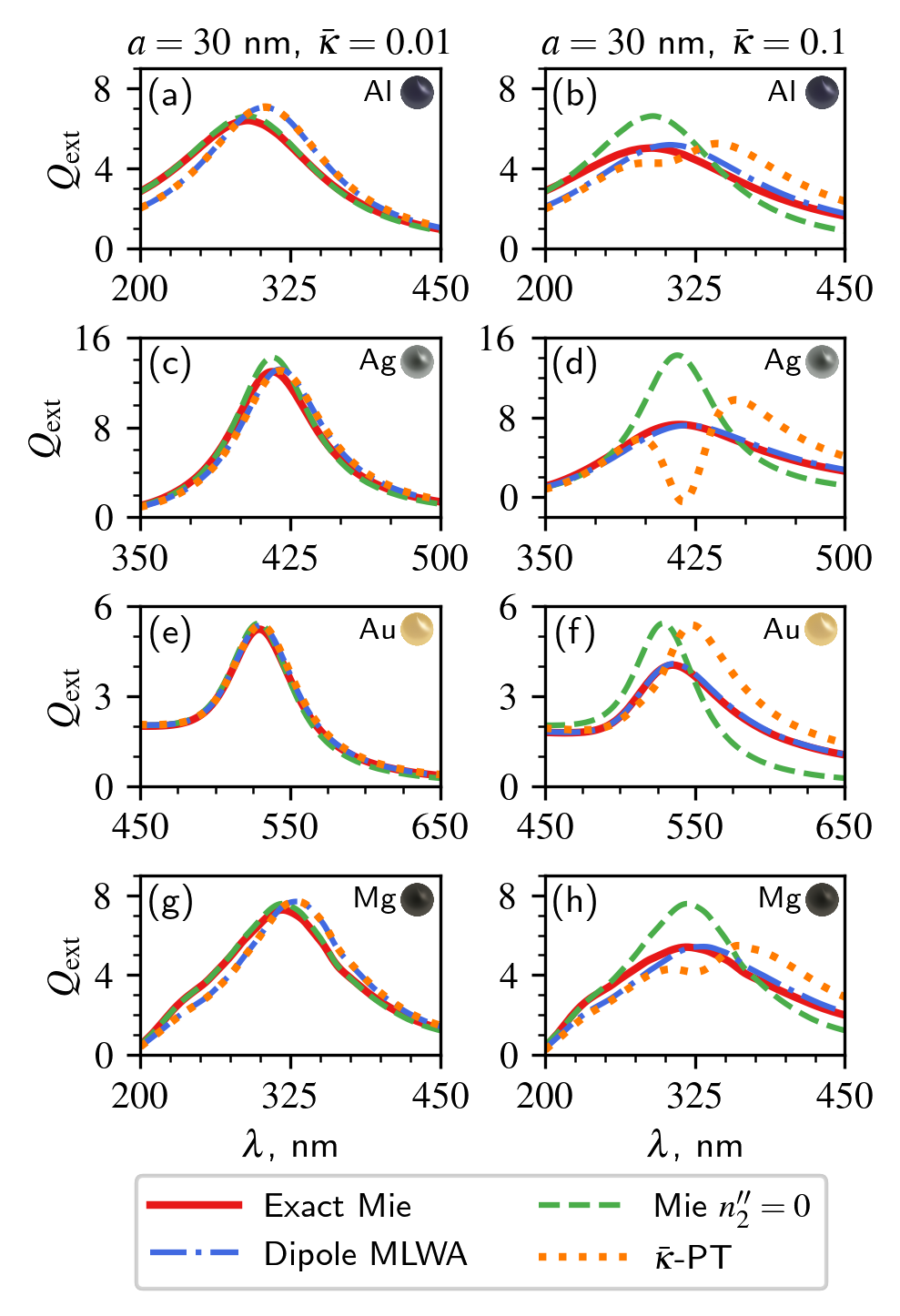}
        \caption{$a=30$~nm}
        \label{fig:mlwa_ext_3}
    \end{subfigure}
    \begin{subfigure}{0.35\textwidth}
        \centering
        \includegraphics[width=\linewidth]{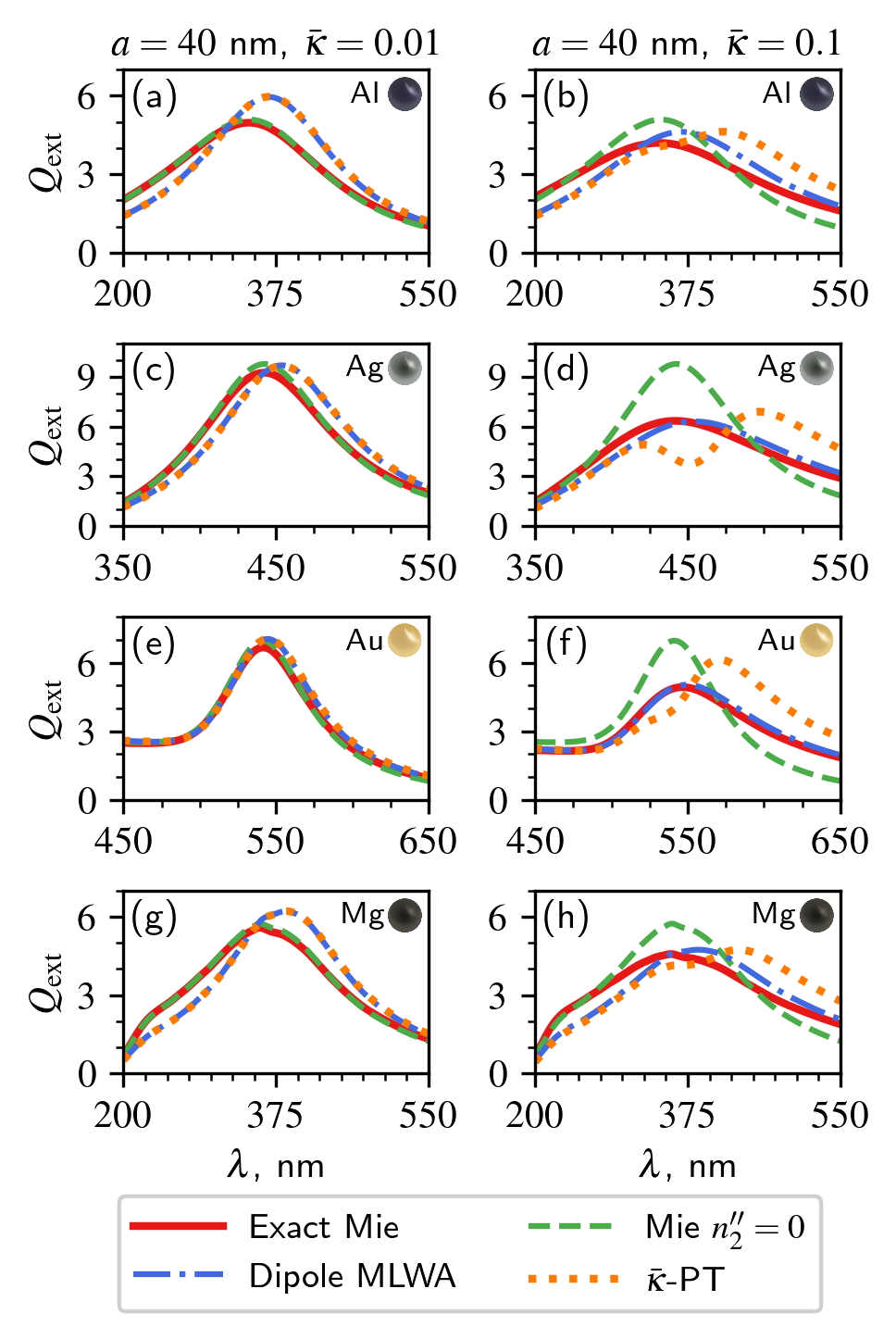}
        \caption{$a=40$~nm}
        \label{fig:mlwa_ext_4}
    \end{subfigure}
    \hfill
    \begin{subfigure}{0.35\textwidth}
        \centering
        \includegraphics[width=\linewidth]{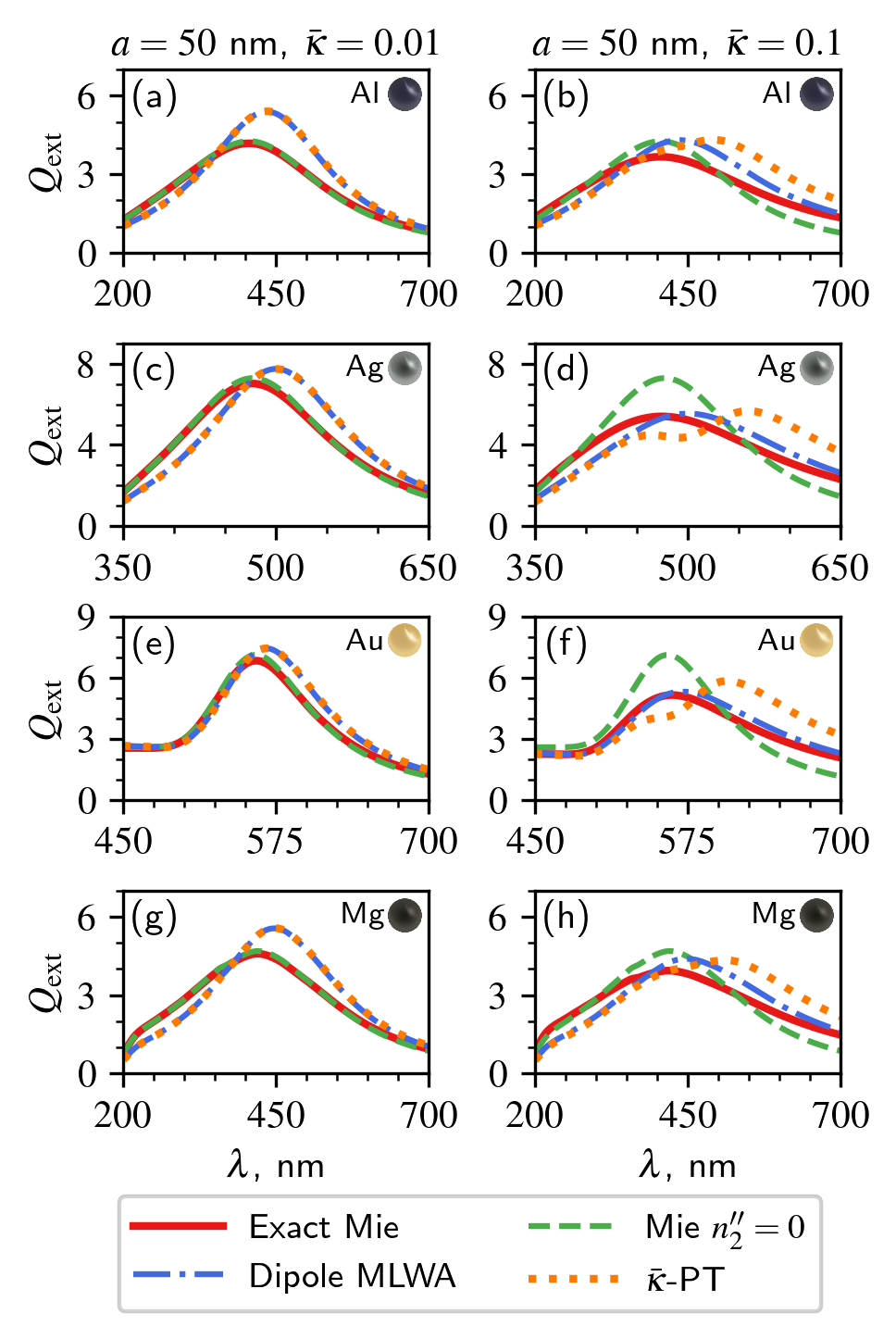}
        \caption{$a=50$~nm}
        \label{fig:mlwa_ext_5}
    \end{subfigure}
    \begin{subfigure}{0.35\textwidth}
        \centering
        \includegraphics[width=\linewidth]{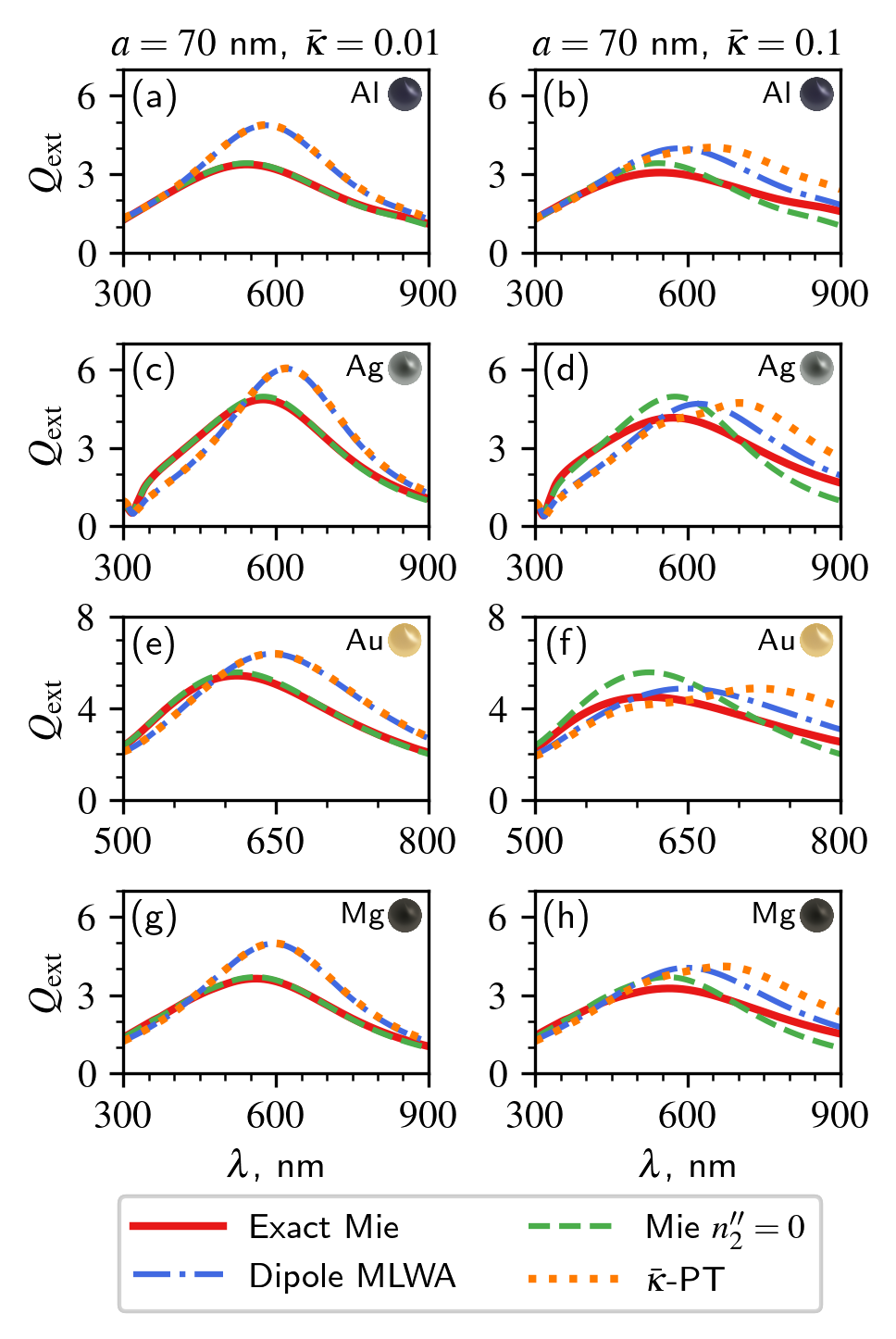}
        \caption{$a=70$~nm}
        \label{fig:mlwa_ext_7}
    \end{subfigure}
    \caption{
    Extinction spectra, $Q_{\rm ext}$, calculated with the exact Mie theory, the Mie theory for non-absorbing media ($n_2''=0$), the dipole MLWA, and the $\bar\kappa$-PT for spherical nanoparticles of Al, Ag, Au and Mg with 
    (I) $a=40$~nm, (II) $a=50$~nm, and (III) $a=70$~nm.
    Absorbing host media have the real part of complex refractive index being that of water ($n_2'=1.33$) and different values of imaginary part:
    $\bar\kappa=0.01$ ($\bar\kappa=0.1$) in the left (right) panel for any given $a$.}
    \label{fig:mlwa_ext_combined}
\end{figure}

\begin{figure}[!hbt]
\centering
    \begin{subfigure}{0.378\textwidth}
        \centering
        \includegraphics[width=\linewidth]{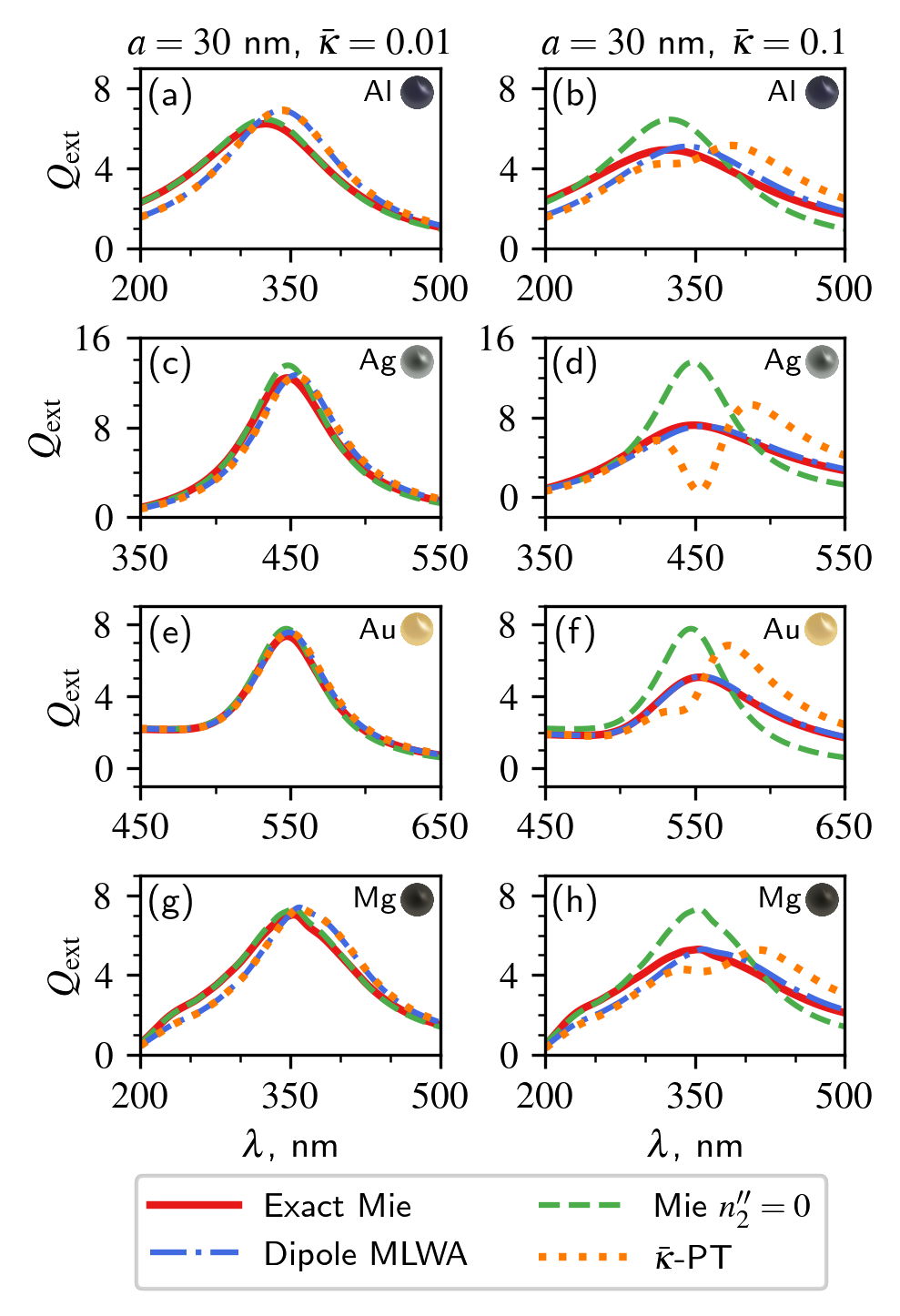}
        \caption{$a=30$~nm}
        \label{fig:mlwa_ext_3_15}
    \end{subfigure}
    \begin{subfigure}{0.37\textwidth}
        \centering
        \includegraphics[width=\linewidth]{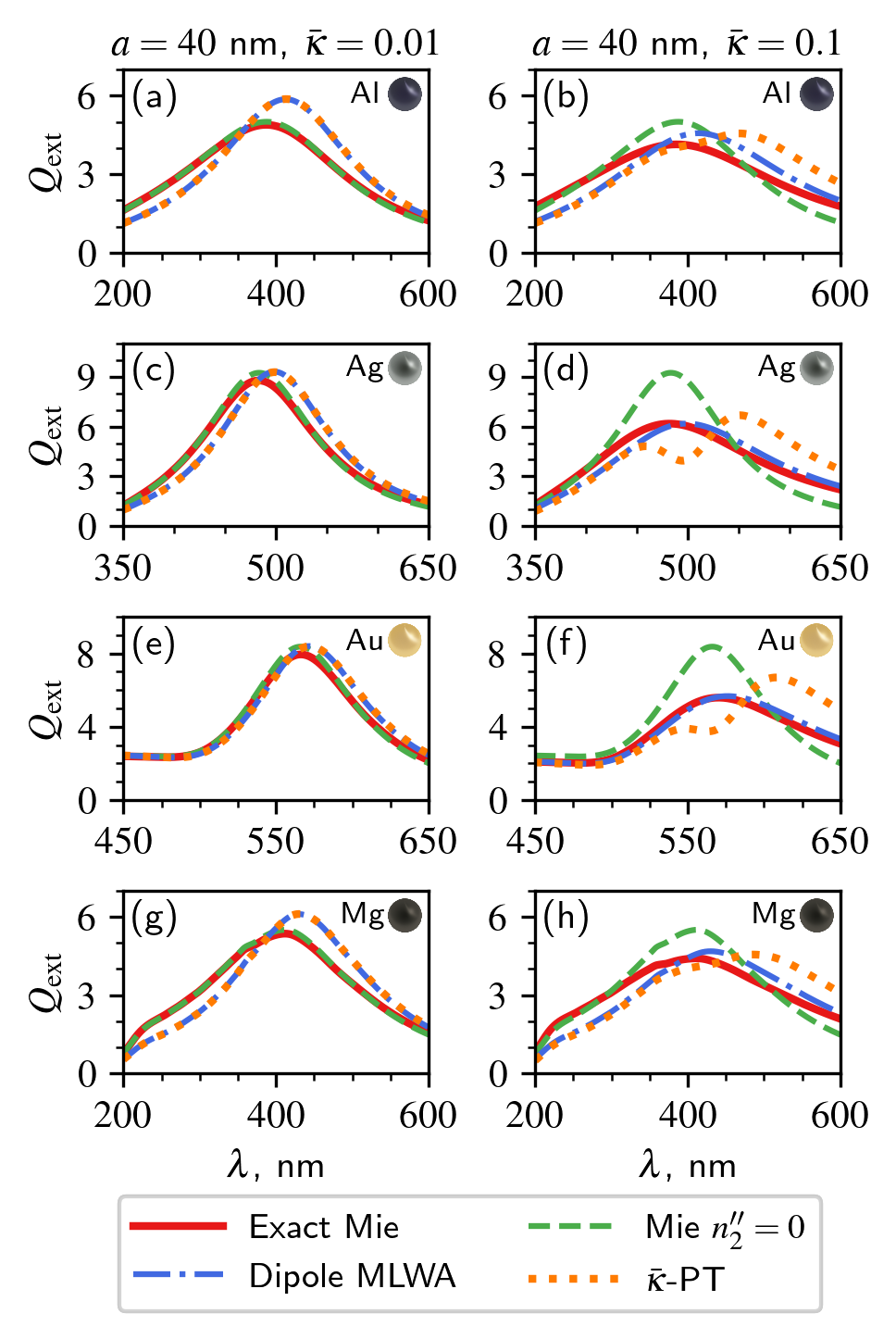}
        \caption{$a=40$~nm}
        \label{fig:mlwa_ext_4_15}
    \end{subfigure}
    \begin{subfigure}{0.37\textwidth}
        \centering
        \includegraphics[width=\linewidth]{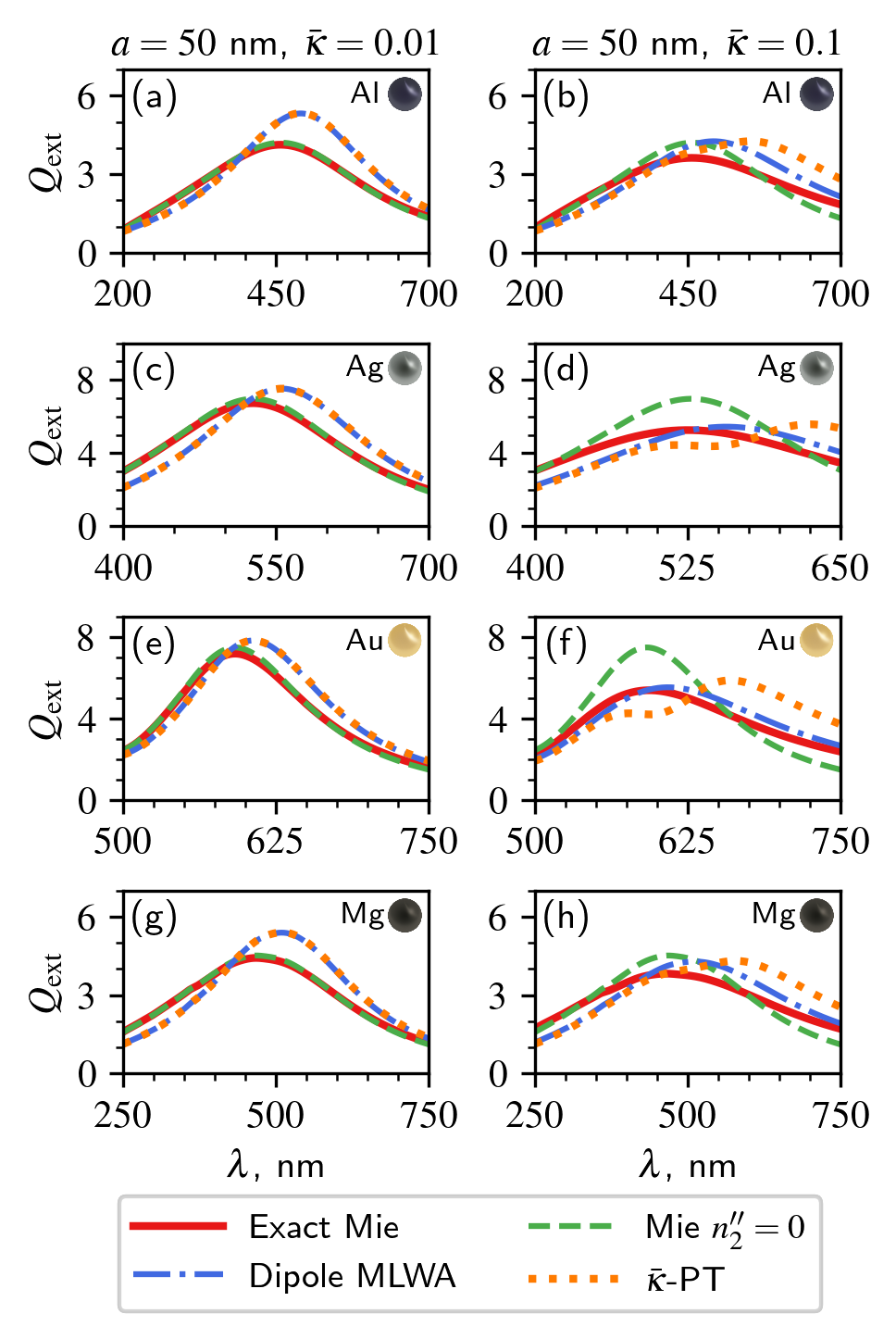}
        \caption{$a=50$~nm}
        \label{fig:mlwa_ext_5_15}
    \end{subfigure}
    \begin{subfigure}{0.37\textwidth}
        \centering
        \includegraphics[width=1.012\linewidth]{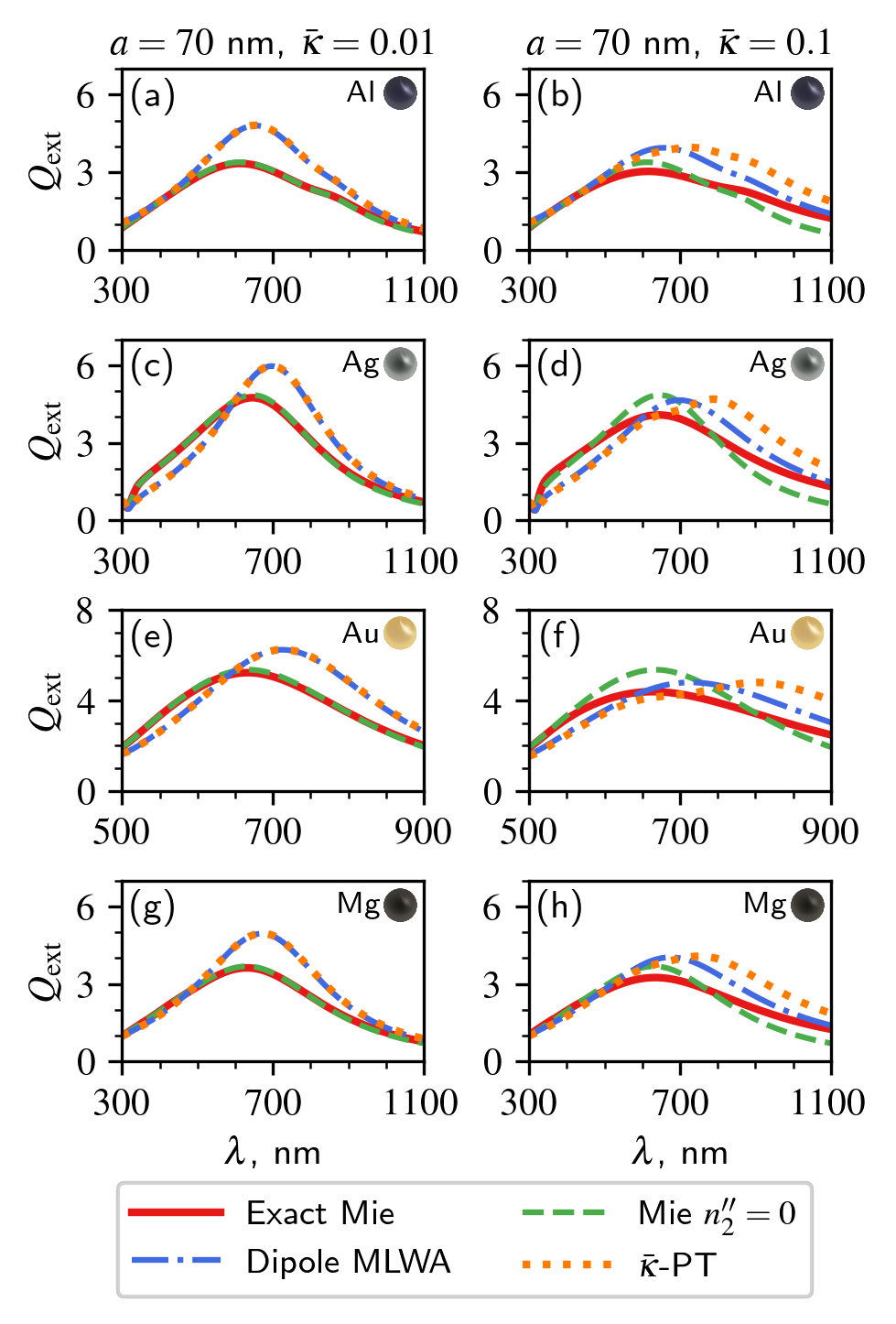}
        \caption{$a=70$~nm}
        \label{fig:mlwa_ext_7_15}
    \end{subfigure}
    \caption{
    Similarly to Fig.~\ref{fig:mlwa_ext_combined}, but in $n_2'=1.5$ host media.
    The left (right) column for any given $a$ is for $\bar\kappa=0.01$ ($\bar\kappa=0.1$).
    }
    \label{fig:mlwa_ext_combined_15}
\end{figure}

\begin{figure}[!ht]
\centering
\begin{subfigure}{0.403\textwidth}
  \centering
  \includegraphics[width=\linewidth]{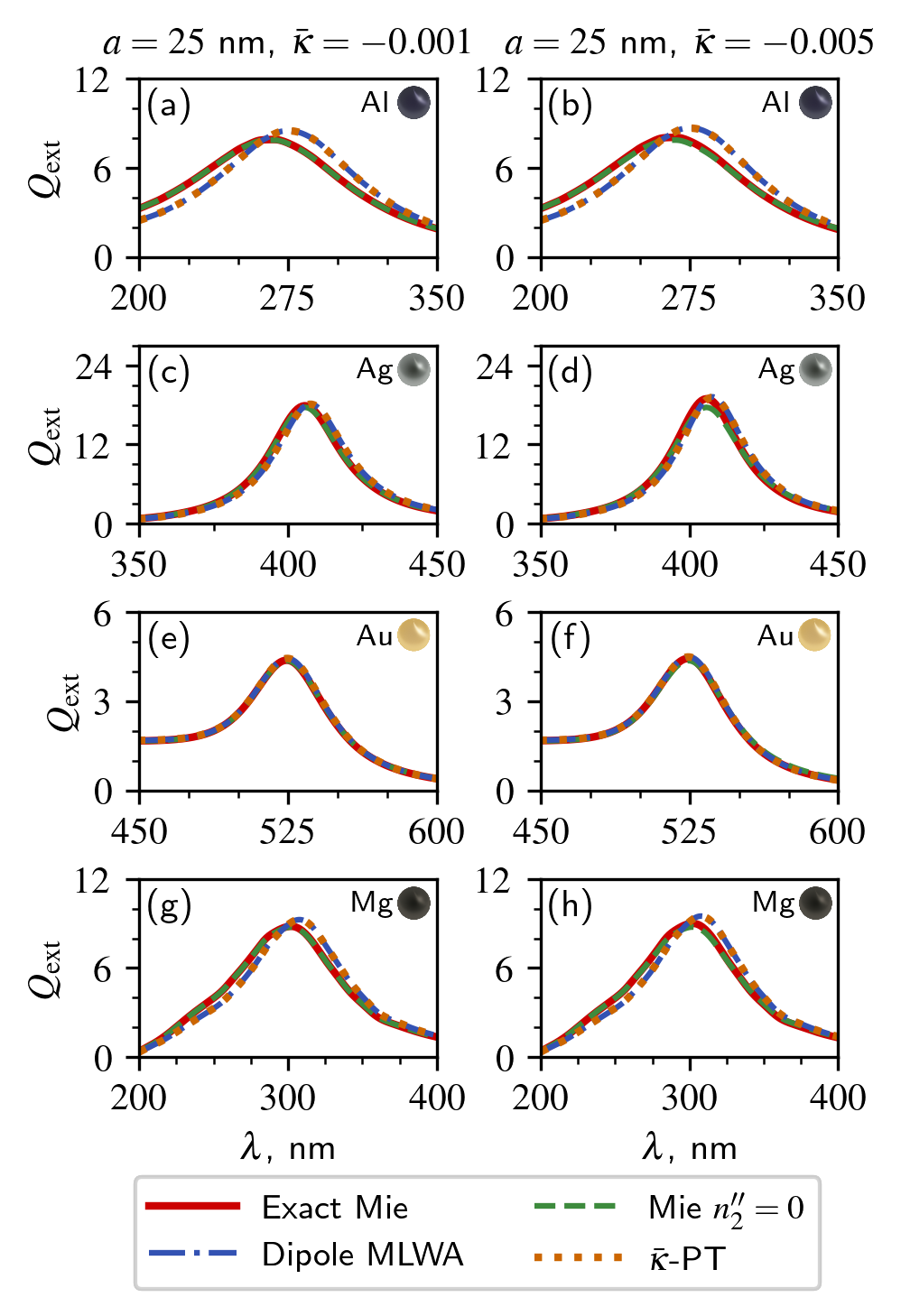}
  \caption{$a=25$~nm}
  \label{fig:mlwa_ext_4gain}
\end{subfigure}
\hspace{30pt}
\begin{subfigure}{0.4\textwidth}
  \centering
  \includegraphics[width=\linewidth]{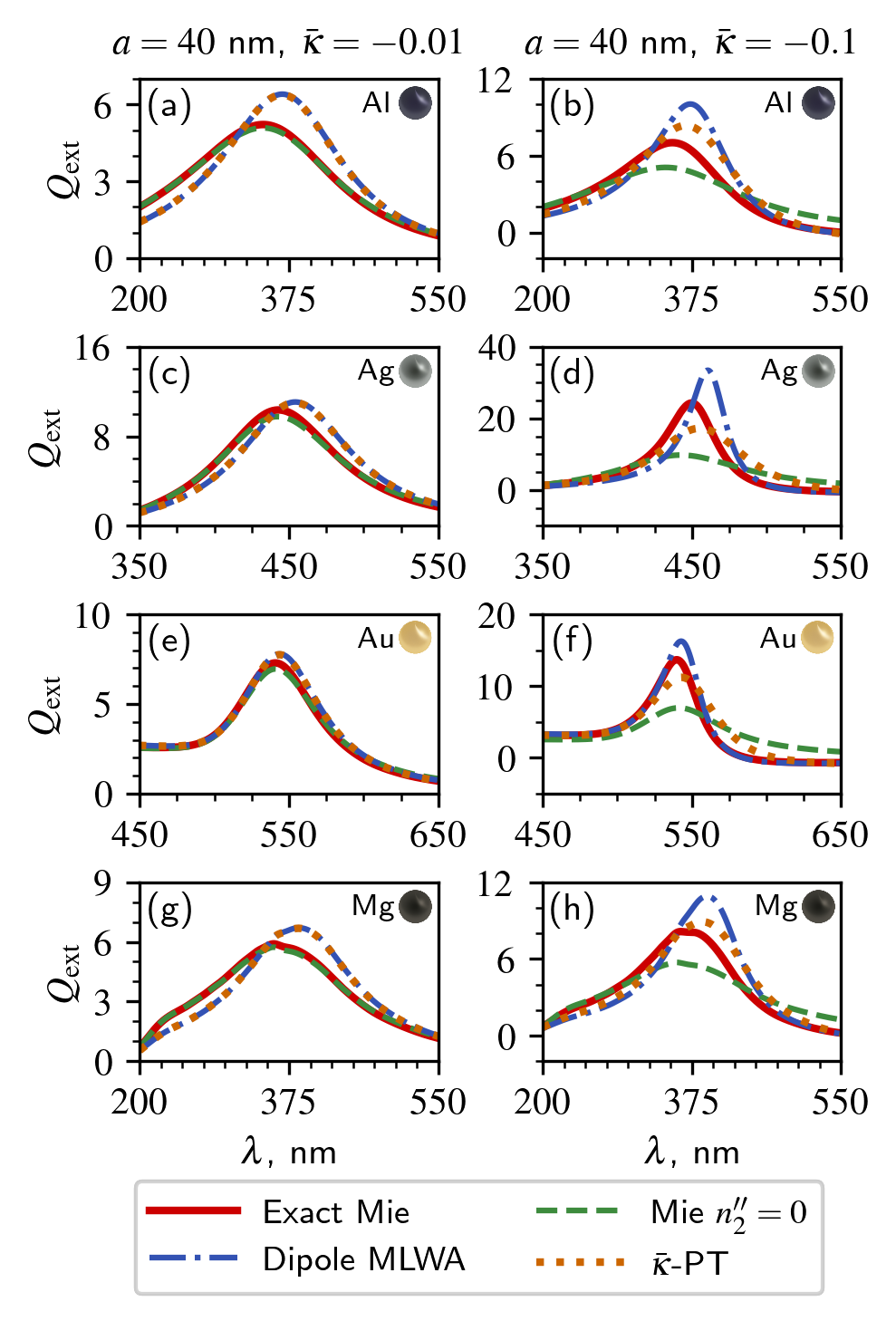}
  \caption{$a=40$~nm}
  \label{fig:mlwa_ext_4gain_si}
\end{subfigure}
\caption{Extinction spectra, $Q_{\rm ext}$, for spherical nanoparticles from Al, Ag, Au and Mg with (I) $a=25$~nm, (II) $a=40$~nm in the gain host media ($n_2'=1.33$).
The left (right) column is for any given $\bar\kappa$, calculated with the exact Mie theory (solid red line), the Mie theory approximation with non-absorbing host with $n_2''=0$ (dashed green line), the dipole MLWA (blue dot-dashed line), and by small $\bar\kappa$ perturbation theory (dotted orange line). Note narrowing of plasmon resonances and increasing of their amplitude with increasing host gain.}
\label{fig:mlwa_ext_gain}
\end{figure}

\begin{figure}[!hbt]
\centering
    \begin{subfigure}{0.4\textwidth}
        \centering
        \includegraphics[width=\linewidth]{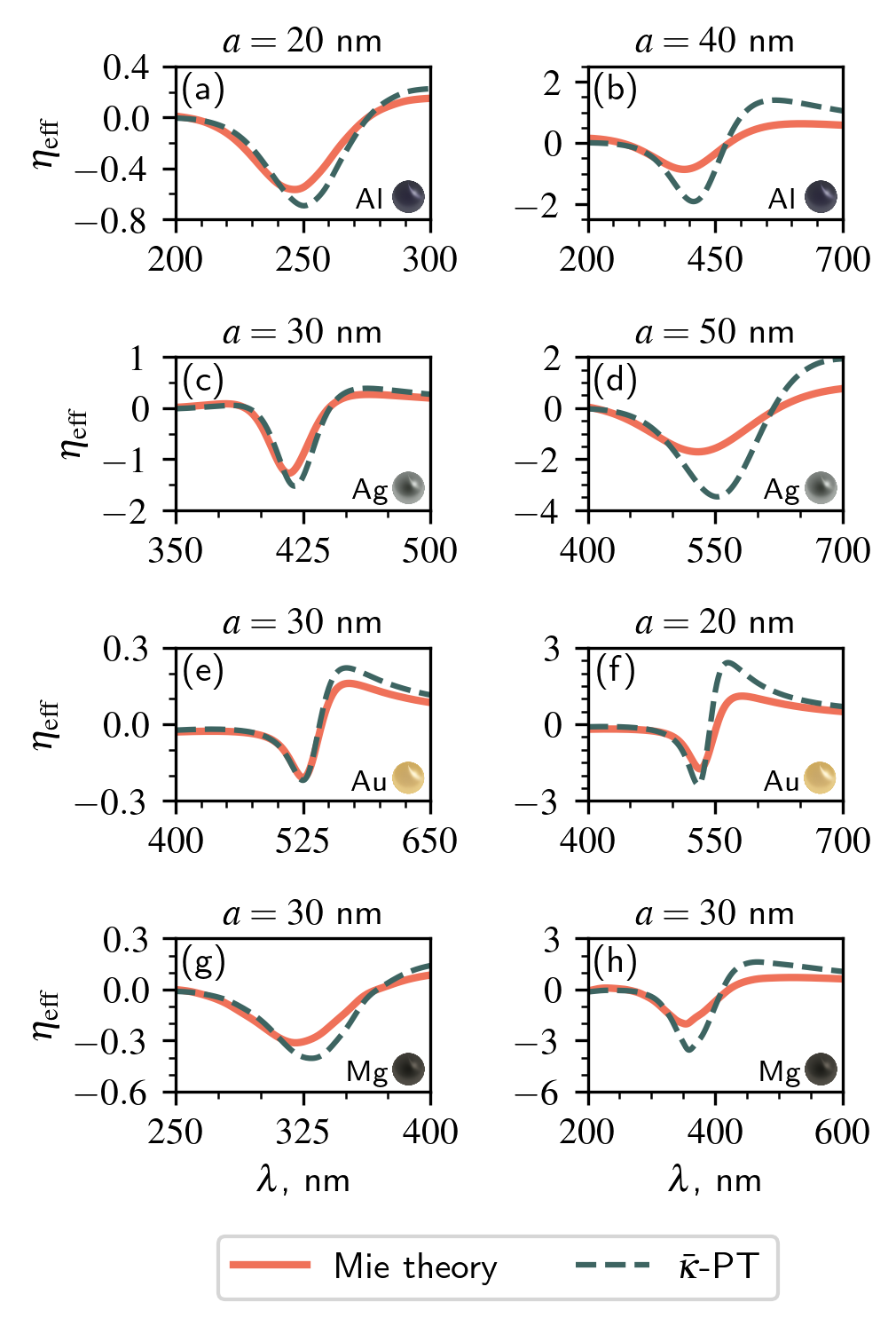}
        \caption{absorbing host media}
        \label{fig:mlwa_ef_si}
    \end{subfigure}
    \begin{subfigure}{0.4\textwidth}
        \centering
        \includegraphics[width=1.012\linewidth]{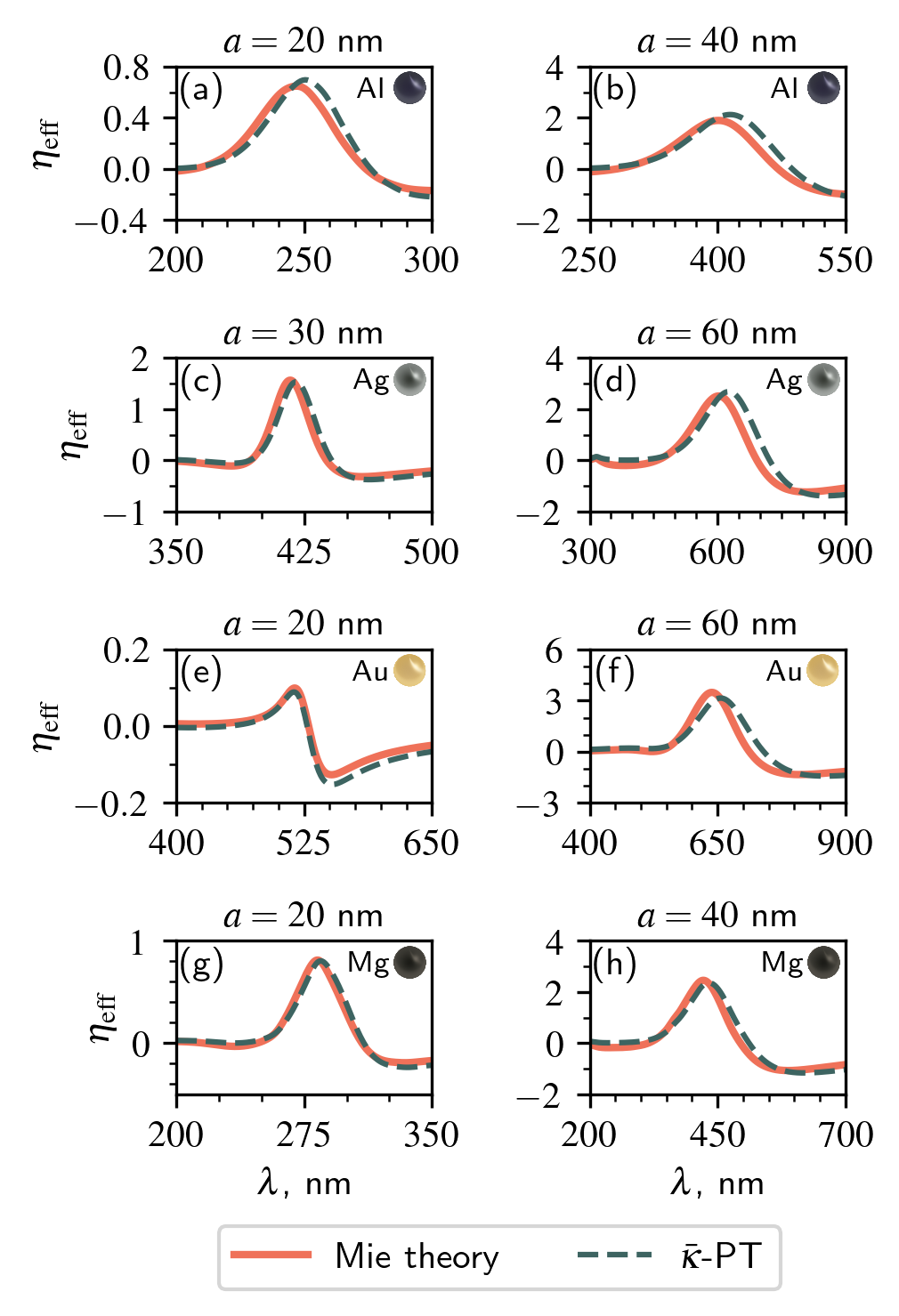}
        \caption{gain host medium}
        \label{fig:mlwa_gain_si}
    \end{subfigure}
    \caption{
    The effect $\eta_{\rm eff}$ of the host absorption on $Q_{\rm ext}$ in the Mie theory (full red line) and the $\bar\kappa$-PT (dashed dark green line) for Al, Ag, Au and Mg materials with different radii $a$ embedded in host medium with the left (right) column for any host medium $n_2'=1.33$ ($n_2'=1.5$) with 
    (I) $\bar\kappa=0.01$ ($\bar\kappa=0.1$) and (II) $\bar\kappa=-0.01$ ($\bar\kappa=-0.1$).
    The figure supplements Fig.~7  of the main text, with the latter showing the results for $\bar\kappa=\pm 0.001$.}
    \label{fig:mlwa_ef_gain_si}
\end{figure}

%%%%%%%%%%%
\begin{figure}[!hbt]
\centering
\includegraphics[width=3.1in]{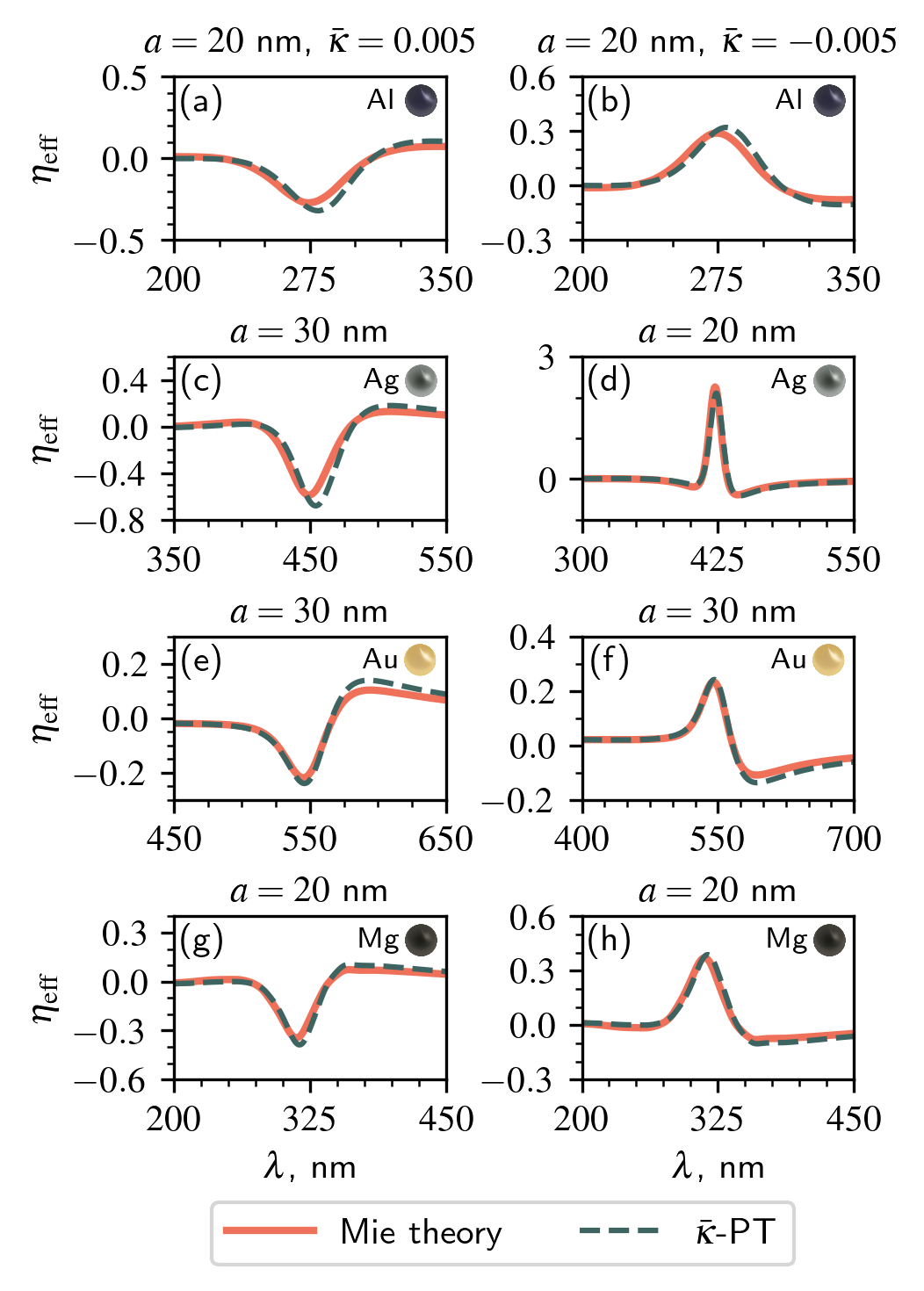}
\caption{The effect $\eta_{\rm eff}$ of the host absorption (left column) and the gain host medium (right column) on $Q_{\rm ext}$ in the Mie theory (solid red line) and the $\bar\kappa$-PT (dashed dark green line) for Al, Ag, Au and Mg materials with different radii $a$ embedded in the host medium with
$n_2'=1.5$ and $\bar\kappa=0.005$ (left column) and
$\bar\kappa=-0.005$ (right column).
The figure supplements Fig.~\ref{fig:mlwa_ef_gain_si} and Fig.~7 of the main text.}
\label{fig:mlwa_ef_gain}
\end{figure}
%%%%%%%%%%%%%%%%%%%%%%%%%%%%

\newpage

\section{Complex pole of the dipole contribution within the dipole MLWA}
\lb{sc:mlwapcz}
%%%%%%%%%%%%%%%%%%%%%%%%%%%%%%%%%%%%%%%%%%%%%%%%%%%%
As stated in the main text, the exceptional feature of the MLWA dipole contribution is that one can determine analytically an exact position of the complex pole of the Mie coefficient $a_1$ at
\bea
\veps_{E1} = -2 \times \fr{1+3x^2/5+ix^3/3}{1-3x^2/5-2ix^3/3},
\lb{sivepsp}
\eea
which corrects Eq.~(6) of ref.~\citenum{SI_Zhang2022b}.
The proof of that $\veps_{E1}$ yields complex zero of the denominator of $a_1$ is relegated to Section~\rf{app:proofzero}.
In the limit of small $x$ one can expand the denominator of $\veps_{E1}$ as $\sim 1+3x^2/5+2ix^3/3$, whereby Eq.~(\rf{sivepsp}) reduces to the familiar classical result (cf. Eq.~(12.13) of ref.~\citenum{SI_Bohren1998})
\begin{equation}
\veps_{E1} \sim \veps_{\rm BH} \approx -2-\frac{12 x^2}{5}\qquad (|x|\ll 1).
\label{sivepsz}
\end{equation}
%%%%%%%%%%%%%%
\begin{figure}[!ht]
\centering
\includegraphics[width=0.65\linewidth]{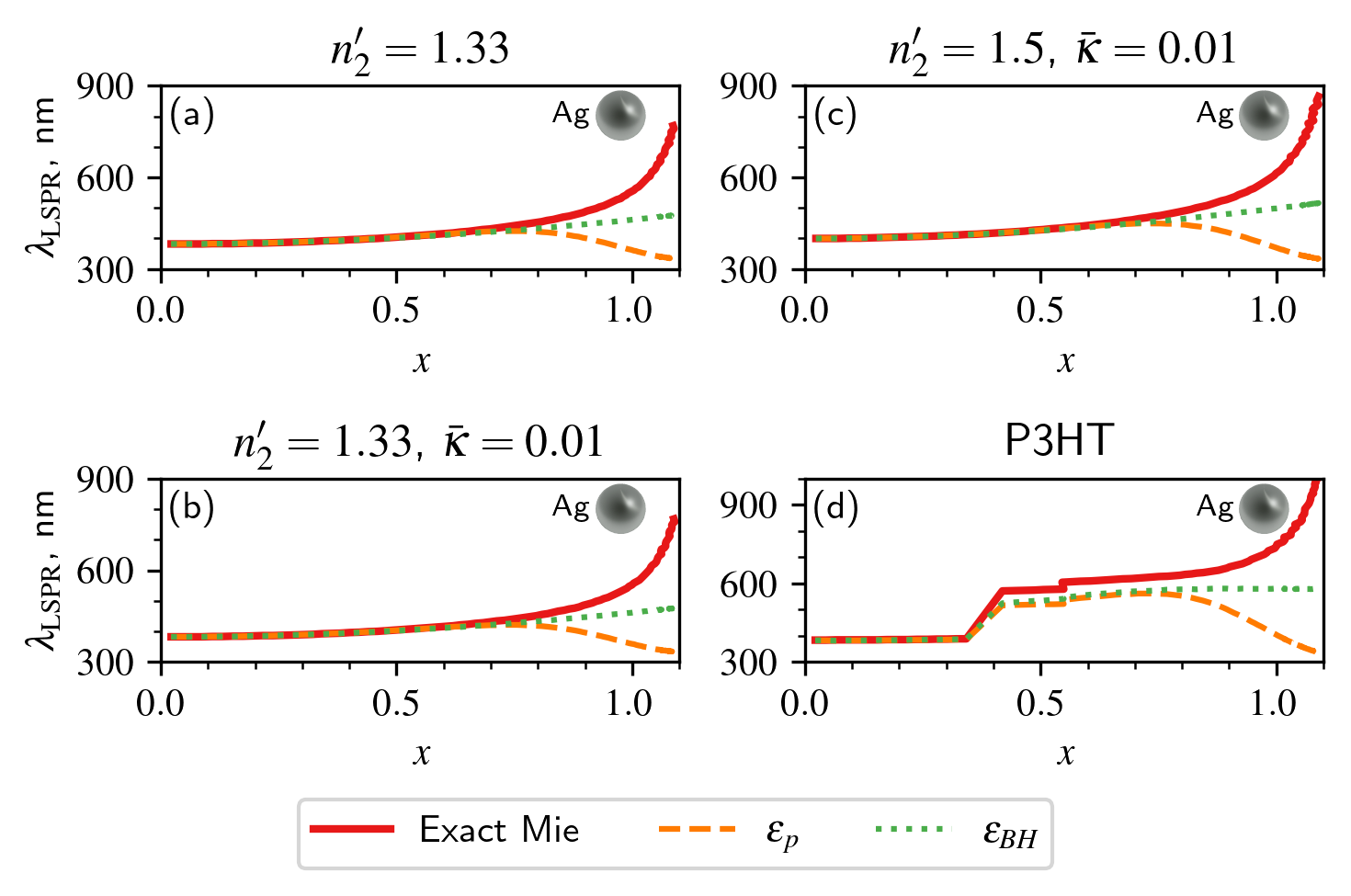}
\caption{The LSPR positions predicted by the MLWA (Eq.~(\rf{sivepsp}) and Eq.~(\rf{sivepsz})) relative to the exact Mie theory (Eq.~~(1)) as functions of size parameter $x$.}
\label{fig:mlwa_mat}
\end{figure}

For a nonabsorbing host, the amended Fr\"ohlich condition (Eq.~(\ref{sivepsz})) has been used to explain the observed initial size-dependent red shift of the dipole localized surface plasmon resonance (LSPR)~\cite{SI_Bohren1998}.
Indeed $\veps_{\rm BH}$ becomes more negative with increasing $x$, which for common Drude metals means that the LSPR is shifted to longer
wavelengths.
Surprisingly, the exact result (Eq.~~(\rf{sivepsp})) only marginally improves the amended Fr\"ohlich condition (Eq.~(\ref{sivepsz})).
Actually the latter works better for $x\sim 1$.

In order to use any of Eq.~(\rf{sivepsp}) or~(\ref{sivepsz}) in practice, one has to notice that each of them does not define an isolated pole but rather a continuous line of poles, one for each given value of $x$.
This is down to small $x$ approximation, because the Mie coefficients in exact Mie theory have only discrete poles.
Writing the relative dielectric function $\veps$ formally also with functional dependence on $x$, one can use Eq.~(\rf{sivepsp}) and~(\ref{sivepsz}) for $|x| \lesssim 1$ to determine the absolute value of the difference $|\veps(x) -\veps_{E1}(x)|$.
The minimum of $|\veps(x) -\veps_{E1}(x)|$ at some $x=x_r$ is expected to provide the dipole LSPR position.
The resulting value of Im $(\veps_{E1}-\eps)$ at $x=x_r$ then determines the resonance linewidth (i.e. the FWHM).
The latter presumes that at the absolute minimum $x=x_r$ of the difference of $|\veps(x) -\veps_{E1}(x)|$ it is justified to apply a pole approximation.
More rigorously, on denoting by $D$ the denominator of $a_1$ in Eq.~(3), one can approximate $D$ around its zero as $D(x) \approx \veps(x)-\veps_{E1}(x_r)$. 
Consequently, near the complex zero of $D$ one has
\bg
a_1\approx - \frac{2i [\veps(x)-1]x^3/3} {\veps(x) -\veps_{E1}(x_r)}\cdot
\lb{mlwa1z}
\eg

Obviously, the complex pole of $a_1$ is a point of maximum of $a_1$ (i.e. dipole LSPR).
A distance from the real axis (i.e. Im~$(\veps_{E1}-\eps))$ determines the resonance linewidth (i.e. the full width at half maximum (FWHM)). 
Had $\veps$ been a real quantity, then the resonance FWHM would be determined simply by Im~$\veps_{E1}$, i.e. the imaginary part of the complex pole as in the conventional scattering theory.
Nonetheless it is difficult to obtain close analytic formulas for the FWHM.

\section{Proof of that $\veps_{E1}$ is the complex zero of the denominator of $a_1$}
\lb{app:proofzero}
%%%%%%%%%%%%%%%%%%%%%%%%%%%%%%%%%%%%%%%%%%%%%%%%%%%%%%
Proof of that $\veps_{E1}$ given by Eq.~(\rf{sivepsp}) is an exact zero of the denominator of $a_1$ in the MLWA is as follows. 
On using Eq.~(\rf{sivepsp}), and on denoting $d=1-3x^2/5-2ix^3/3$ the denominator of $\veps_{E1}$ in Eq.~(\rf{sivepsp}), one finds
\bea
\veps_{E1} +2 &=& - \fr{2}{d} \, (1+3x^2/5+ix^3/3-1+3x^2/5+2ix^3/3)
\nn\\
&=& - \fr{6}{d} \, (2x^2/5+ix^3/3),
\nn\\
3(\veps_{E1}-2) &=& - \fr{6}{d} \, (1+3x^2/5+ix^3/3+1-3x^2/5-2ix^3/3)
\nn\\
&=&  - \fr{6}{d} \, (2-ix^3/3),
\nn\\
2(\veps_{E1}-1) &=& - \fr{2}{d} \,(2+6x^2/5+2ix^3/3+1-3x^2/5-2ix^3/3)
\nn\\
&=& - \fr{6}{d} \,(1+x^2/5).
\eea
On denoting $D$ the denominator of $a_1$ in the MLWA (Eq.~(3)), then at $\veps =\veps_{E1}$
\bea
D &=&
- \fr{6}{d} \, \left[2x^2/5+ix^3/3  - (x^2/5)(2-ix^3/3)
-i (x^3/3)(1+x^2/5) \right]
\nn\\
&=&
- \fr{6}{d} \, \left[ i ((x^2/5)x^3/3)-i ((x^3/3)x^2/5) \right]\equiv 0.
\eea

\section{The imaginary part of $\veps_{E1}$ is negative}
\lb{sc:imep}
%%%%%%%%%%%%%%%%%%%%%%%%%%%%%%%%%%%%%%%%%%%%%%%%%%%%%
Here we show that the imaginary part of $\veps_{E1}$ given by Eq.~(\rf{sivepsp}) is negative for real $x$.
One finds
\bea
\veps_{E1} &=& -\fr{2}{(1-3x^2/5)^2+(2x^3/3)^2}
\nn\\
&& \times
(1+3x^2/5+ix^3/3)(1-3x^2/5+2ix^3/3)
\nn\\
&=& -\fr{2}{(1-3x^2/5)^2+(2x^3/3)^2}
\nn\\
&& \times
\left[ 1-(3x^2/5)^2 - 2 (x^3/3)^2
+ i(x^3/3)(1+x^2/5)\right].
\lb{vepspa}
\eea
Hence
\bg
\mb{Im } \veps_{E1} = -\fr{(2/3)x^3(1+x^2/5)}{(1-3x^2/5)^2+(2x^3/3)^2}
< 0.
\lb{vepspi}
\eg

\section{The derivative of $dT_{E1}$}%$\bar\kappa$
\lb{sc:1der}
%%%%%%%%%%%%%%%%%%%%%%%%%%%%%%%%%%%%%
Taking into account $\kappa$-dependence of 
both $\veps$ (cf. Eq.~(6) of the main text) and of the size parameter $x=2\pi a n_2/\lambda$ through its explicit dependence on the complex refractive index 
$n_2=n_2'(1+i\kappa)$, the derivative $dT_{E1}/d\bar\kappa$ is calculated as
\bg
\fr{dT_{E1} }{d\bar\kappa}=\fr{\pa T_{E1} }{\pa \veps}\fr{d\veps }{d\bar\kappa}
+\fr{\pa T_{E1} }{\pa x}\fr{dx }{d\bar\kappa},
\lb{dTdef}
\eg
where (cf. Eq.~(6))
\bea
\fr{d\veps}{d\bar\kappa} &=& -\fr{2i\veps}{(1+i\bar\kappa)^3}
\to -2i\veps_t \qquad (\bar\kappa\to 0),
\nn\\
\fr{dx}{d\bar\kappa}  &=& i n_2' x_0=i x' \qquad (\bar\kappa\to 0).
\eea
For the future convenience, we have introduced the size parameter in vacuum host, $x_0=2\pi a /\lambda$, leading to $x=n_2 x_0 = n_2' x_0 (1+i\bar\kappa)$.
On recalling Eq.~(3),
\bea
T_{E1}  &=& \frac{2i (\veps-1)x^3/3} {\veps +2 -3(\veps-2)x^2/5 -2i (\veps-1)x^3/3} 
\cdot
\lb{TE10}
\eea
Now
\bea
\fr{\pa T_{E1} }{\pa x} = \frac{\fr{\pa u}{\pa x}D-u\fr{\pa D}{\pa x}}{D^2},
\eea
where
\bea
u &=& \frac{2}{3}ix^3 (\veps-1),
\nn\\
\fr{\pa u}{\pa x} &=& 2ix^2 (\veps-1),
\nn\\
\fr{\pa u}{\pa \veps}  &=& \frac{2}{3}ix^3,
\nn
\eea

\bea
D &=& \veps +2 -\fr{3}{5}x^2 (\veps-2) -\frac{2}{3}ix^3 (\veps-1),
\lb{Depst}
\\
\fr{\pa D}{\pa x} &=& -\fr{6}{5}x(\veps-2)- 2ix^2 (\veps-1),
\nn\\
\fr{\pa D}{\pa \veps} &=& 1-\frac{3}{5}x^2-\frac{2}{3}ix^3,
\nn
\eea

\bea
u\fr{\pa D}{\pa x}  &=& \frac{2}{3}ix^3 (\veps-1) \left[-\fr{6}{5}x(\veps-2)- 2ix^2 (\veps-1) \right]
\nn\\
&=& -\frac{4}{5}ix^4 (\veps-1)(\veps-2) + \frac{4}{3}x^5 (\veps-1)^2,
\nn
\eea

\bea
\fr{\pa u}{\pa x}D &=& 2ix^2 (\veps-1) \left[\veps +2 -\fr{3}{5}x^2 (\veps-2) -\frac{2}{3}ix^3 (\veps-1)\right]
\nn\\
&=& 2ix^2 (\veps-1)(\veps +2) - \frac{6}{5}ix^4 (\veps-1)(\veps-2) + \frac{4}{3}x^5 (\veps-1)^2,
\nn
\eea

\bea
\fr{\pa u}{\pa x}D-u\fr{\pa D}{\pa x} &=& 2ix^2 (\veps-1)(\veps +2) - \frac{6}{5}ix^4 (\veps-1)(\veps-2) + \frac{4}{3}x^5 (\veps-1)^2
\nn\\
&+& \frac{4}{5}ix^4 (\veps-1)(\veps-2) - \frac{4}{3}x^5 (\veps-1)^2
\nn\\
&=& 2ix^2 (\veps+2)(\veps-1) - \frac{2}{5}ix^4(\veps-1)(\veps-2),
\nn
\eea

\bea
\fr{\pa T_{E1} }{\pa x} =2ix^2 (\veps - 1)\,  \frac{\veps+2 -\frac{1}{5}x^2 (\veps-2)}{D^2}\cdot
\nn
\eea
Taking the limit $\bar\kappa\to 0$
\bea
\lim_{\bar\kappa\to 0}
\fr{\pa T_{E1} }{\pa x}\fr{dx}{d\bar\kappa} =- 2(x')^3 (\veps_t - 1)\,  \frac{\veps_t +2 -\frac{1}{5}(x')^2 (\veps_t-2)}{D^2(\veps_t,x')},
\lb{dTxl}
\eea
where the functional dependence $D(\veps_t,x')$ indicates that $D$ in Eq.~(\rf{Depst}) is a function of $\veps_t$ and $x'$ in the limit case.

%%%%%%%%%%%%%%%%%%%%%%%%%
Analogously, 
\bea
\fr{dT_{E1} }{\pa \veps} = \frac{\fr{\pa u}{\pa \veps}D - u\fr{\pa D}{\pa \veps}}{D^2},
\eea

\bea
u\fr{\pa D}{\pa \veps}  &=& \frac{2}{3}ix^3 (\veps-1) \left[1-\frac{3}{5}x^2-\frac{2}{3}ix^3\right]
\nn\\
&=& \frac{2}{3}ix^3(\veps-1) - \frac{2}{5}ix^5(\veps-1) + \frac{4}{9}x^6(\veps-1),
\eea

\bea
\fr{\pa u}{\pa \veps} D &=& \frac{2}{3}ix^3 \left[\veps +2 -\frac{3}{5}x^2(\veps-2) -\frac{2}{3}ix^3 (\veps-1)\right]
\nn\\
&=& \frac{2}{3}ix^3 (\veps+2) -\fr{2}{5}ix^5 (\veps-2) + \frac{4}{9}x^6 (\veps-1),
\eea

\bea
\fr{\pa u}{\pa \veps}D - u\fr{\pa D}{\pa \veps} &=& \frac{2}{3}ix^3 (\veps+2) -\fr{2}{5}ix^5 (\veps-2) + \frac{4}{9}x^6 (\veps-1)
\nn\\
&-& \frac{2}{3}ix^3(\veps-1) + \frac{2}{5}ix^5(\veps-1) - \frac{4}{9}x^6(\veps-1)
\nn\\
&=& 2ix^3 + \fr{2}{5}ix^5= 2ix^3 \left( 1 + \fr{1}{5}x^3\right),
\eea

\bg
\fr{\pa T_{E1} }{\pa\veps} = 2ix^3 \frac{ (1+x^2/5)}{D^2}\cdot
\eg
Taking the limit $\bar\kappa\to 0$
\bea
\lim_{\bar\kappa\to 0}
\fr{\pa T_{E1} }{\pa\veps} \fr{d\veps}{d\bar\kappa} = 
4 \veps_t (x')^3  \frac{1+(x')^2/5}{D^2(\veps_t,x')}\cdot
\lb{dTel}
\eea

%%%%%%%%%%%%%%%%%%%%%%%%%
On assembling the intermediary steps 
(\rf{dTdef}), (\rf{dTxl}), (\rf{dTel}) together,
\bea
\lim_{\bar\kappa\to 0} \fr{dT_{E1} }{d\bar\kappa} &=&
\lim_{\bar\kappa\to 0}  \fr{\pa T_{E1} }{\pa x}\fr{dx }{d\bar\kappa}
+
\lim_{\bar\kappa\to 0} \fr{\pa T_{E1} }{\pa \veps}\fr{d\veps }{d\bar\kappa}
\nn\\
&=& 
- 2(x')^3 (\veps_t - 1)\,  \frac{\veps_t +2 -\frac{1}{5}(x')^2 (\veps_t-2)}{D^2(\veps_t,x')}
+ 4 \veps_t (x')^3  \frac{1+(x')^2/5}{D^2(\veps_t,x')}
\nn\\
&=& \fr{(x')^3}{D^2(\veps_t,x')} \, \left\{4\veps_t [1 + (x')^2/5] - 2  (\veps_t+2)(\veps_t-1) + \fr{2}{5} (x')^2(\veps_t-2)(\veps_t-1) 
\right\}
\nn\\
& = & \fr{(x')^3}{D^2(\veps_t,x')} \, \left\{
4\veps_t - 2  (\veps_t+2)(\veps_t-1)+ (2/5)(x')^2[2\veps_t + (\veps_t-2)(\veps_t-1)]
\right\}
\nn\\
& = & \fr{(x')^3}{D^2(\veps_t,x')} \, \left\{
4\veps_t - 2  (\veps_t^2+\veps_t-2)+ (2/5)(x')^2(2\veps_t + \veps_t^2 -3\veps_t+2)
\right\}
\nn\\
& = & \fr{2(x')^3}{D^2(\veps_t,x')} \, \left\{ 2+ \veps_t-\veps_t^2 + \fr{1}{5}(\veps_t^2-\veps_t +2) (x')^2 
\right\},
\eea
and
\bea
\left. \fr{da_1}{d\bar\kappa}\right|_{\bar\kappa\to 0}  &=&\left. - \fr{dT_{E1}}{d\bar\kappa}\right|_{\bar\kappa\to 0}  
= -\fr{2 (x')^3}{D^2(\veps_t,x')} \, \left\{ 2+ \veps_t-\veps_t^2 + \fr{1}{5}(\veps_t^2-\veps_t +2)  (x')^2 \right\}.
\lb{a1d}
\eea
Eventually, 
\bea
Q_{\rm ext} &=& \fr{C_{\rm ext}}{\pi a^2} \approx \fr{2}{x'}\, \mb{Re}\left\{  \fr{3}{x}\, \left[ a_1(\veps_{t}) + \bar\kappa \left. \fr{da_1}{d\bar\kappa}\right|_{\veps=\veps_{t}} \right]
\right\}
\nn\\
&=& \fr{2}{x'} \mb{Re} \left\{\fr{3}{x} a_1 (\veps_{t}) \right\} -12 n_2' x' \bar\kappa\, \mb{Re}\left[\fr{1}{n_2 D^2(\veps_t,x')} \, \left\{ 2- \veps_t(\veps_t-1) + \fr{1}{5}(\veps_t^2-\veps_t +2) (x')^2\right\} \right],
\nn\\
&&
\lb{siCextl1}
\eea
where we have made use of that $x'=x_0 n_2'$ is a real number and $x'/x=n_2'/n_2$, where $n_2=n_2'+in_2''$ is the complex refractive index.
Equation~(\rf{siCextl1}) defines the first order perturbation expansion (PT) of $Q_{\rm ext}$ in the parameter $\bar\kappa$ in the main text.
We remind here that $\veps_t$ is in general a complex number and, like $n_2$ in the denominator, cannot be taken in front of the Re sign.

%%%%%%%%%%%%%%%%%%
\begin{figure}[!htb]
\centering
\includegraphics[width=0.5\linewidth]{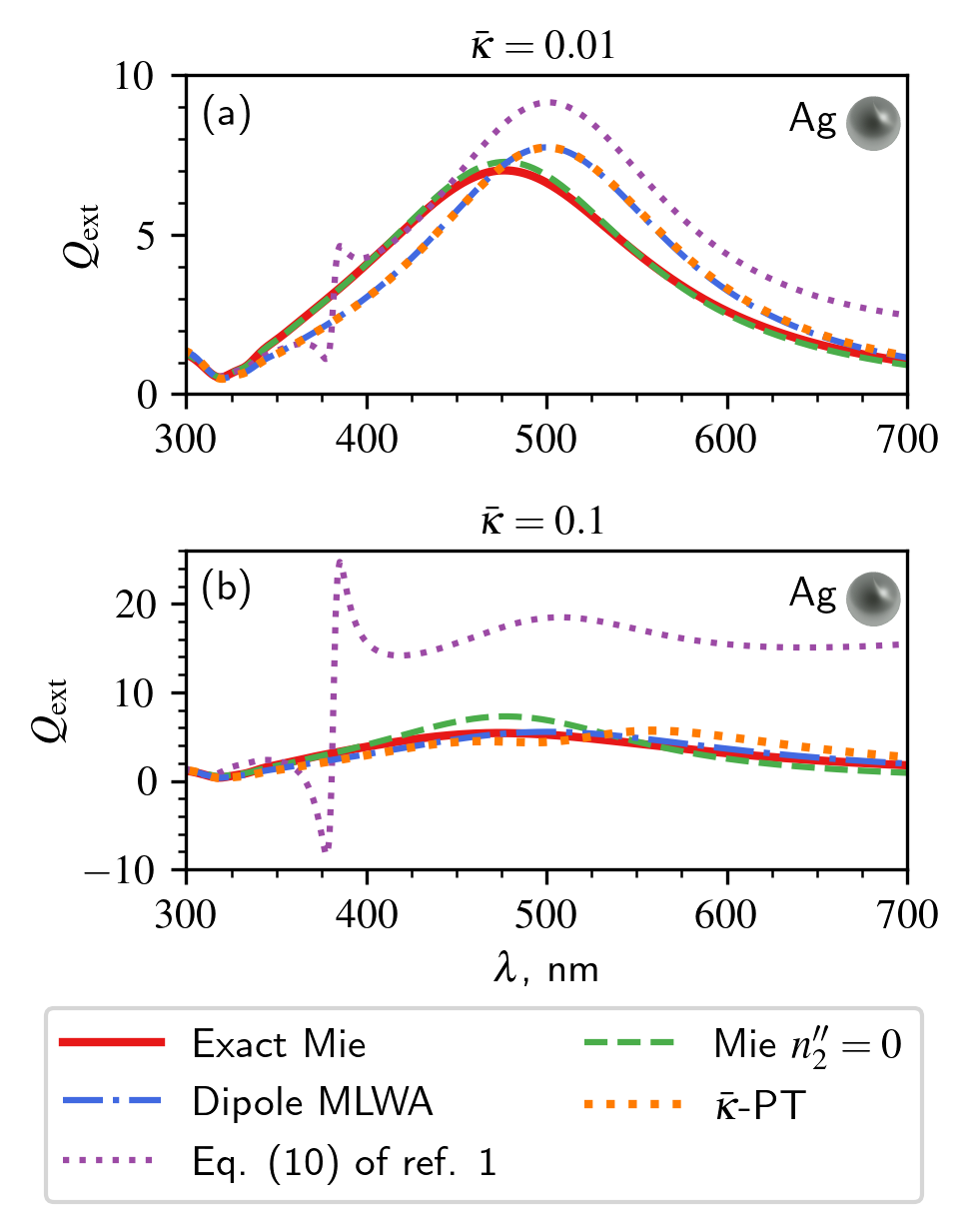}
\caption{Extinction spectra, $Q_{\rm ext}$, for spherical Ag particles with radius $a=50$~nm embedded in host medium with
(a) $n_2=$1.33+0.0133${\rm i}$ and
(b) $n_2=$1.33+0.133${\rm i}$, calculated by Eq.~(1) of the Mie theory (solid red line), 
the Mie theory in nonabsorbing host with $n_2''=0$ (green dashed line), the  dipole MLWA of Eq.~(3) (blue dot-dashed line), and the $\bar\kappa$-PT of our Eq.~(9) (dotted orange line).
Small $\bar\kappa$-PT given by Eq.~(10) of ref.~\citenum{SI_Zhang2022b} is shown in velvet dotted line.}
\label{fig:mlwa_comp}
\end{figure}
%%%%%%%%%%%%
\section{A previous attempt to formulate a perturbation theory (PT)}%  $\bar\kappa$ 
\lb{sec:ppt}
%%%%%%%%%%%%%%%%%%%%%%%%%%%%%%%%%%%%%%%%%%%%%
A first attempt to formulate a PT in terms of $\bar\kappa$ has been performed earlier in ref.~\citenum{SI_Zhang2022b}.
However we could neither reproduce their analytic results nor confirm that they are in any sense reliable.
For the sake of comparison, Fig.~\ref{fig:mlwa_comp} displays a comparison of the respective PT's against the exact Mie theory.
The small $\bar\kappa$ approximation, given earlier by Eq.~(10) of ref.~\citenum{SI_Zhang2022b} and shown in the velvet dotted line, performs worst and fails dramatically for $\bar\kappa=0.1$.
The cause of it is seen in a computational error.

\section{Apparent extinction cross section}
\lb{sec:aextcs}
%%%%%%%%%%%%%%%%%%%%%%%%%%%%%%%%%%%%%%%%%%%%%
In order to accommodate the changes for an {\em absorbing} host, we have to recast 
the extinction cross sections as
\bea
\lefteqn{
\sigma_{ext;p\ell} = -\frac{2\pi (2\ell +1)}{k'}\, \mb{Re }
\left(\fr{1}{k}\fr{i R(x)} {F + D(x) -i R(x)}\right)
}
\nn\\
&=& \frac{2\pi (2\ell +1)}{k'}\, \mb{Im }
\left(\fr{1}{k}\fr{ R(x)} {F + D(x) -i R(x)}\right).
\label{sgtotah}
\eea
Within the dipole (i.e. $\ell=1$) MLWA~\ct{SI_Rasskazov2020a,SI_Rasskazov2021} the above cross-sections take on the following form:
\begin{eqnarray}
\sg_{\rm sca;1} &=& \frac{4\pi}{15 k^2}\, \frac{10\, x^6 \left| \tl\veps_1-1\right|^2}{\left|\tl\veps_1+2-\frac{3}{5} (\tl\veps_1-2) x^2-i \frac{2}{3} (\tl\veps_1-1) x^3\right|^2},
\label{sgscsdm}
\\
\sg_{\rm abs;1} &=& \frac{4\pi}{15 k^2}\, \frac{9\, x^3 \left(x^2+5\right) \mb{Im }(\tl\veps_1)}{\left|
\tl\veps_1+2-\frac{3}{5} (\tl\veps_1-2) x^2-i \frac{2}{3} (\tl\veps_1-1) x^3\right|^2},
\label{sgabsdm}
\\
\sg_{\rm ext;1} &=& \frac{4\pi}{15 k^2}\, \frac{9\, x^3 \left(x^2+5\right) \mb{Im }(\tl\veps_1)+10\, x^6 \left|\tl\veps_1-1\right|^2}
{\left|\tl\veps_1+2-\frac{3}{5} (\tl\veps_1-2) x^2-i \frac{2}{3} (\tl\veps_1-1) x^3\right|^2},
\label{sgtotsdm}
\end{eqnarray}
where $\mb{Im }(\tl\veps_1)$ denotes the imaginary part of $\tl\veps_1$.
The higher order multipole MLWA can be treated similarly~\ct{SI_Rasskazov2021}.

\section{Conventional scattering theory}
\lb{sec:csth}
%%%%%%%%%%%%%%%%%%%%%%%%%%%%%%%%%%%%%%%%%%
In conventional scattering theory, any given angular momentum $\ell$ and polarization $p$ ($p=E$ for electric (or TM) polarization, and $p=M$ for magnetic (or TE) polarization) channel contributes the following partial amount to the resulting scattering, absorption, and extinction cross sections (Eqs.~(2.135-8) of ref.~\citenum{SI_Newton1982}),
\begin{eqnarray}
\sigma_{{\rm sca};p\ell} &= & \frac{2\pi (2\ell +1)}{k^2}\, |T_{p\ell}|^2,
\label{sgsc}
\\
\sigma_{{\rm abs};p\ell} &= & - \frac{2\pi (2\ell +1)}{k^2}\, \left[|T_{p\ell}|^2 +
\mb{Re }(T_{p\ell}) \right],
\label{sgabs}
\\
\sigma_{{\rm ext};p\ell} &= & -\frac{2\pi (2\ell +1)}{k^2}\, \mb{Re }(T_{p\ell}),
\label{sgtot}
\end{eqnarray}
where $k=2\pi/\ld$ is the wavenumber, with $\ld$ being the incident 
wavelength in the \textit{host medium}.

The resulting full cross sections are determined as an infinite sum
\begin{equation}
 \sigma_{\rm sca} = \sum_{p,\ell} \sigma_{{\rm sca};p\ell} \ , \quad 
 \sigma_{\rm abs} = \sum_{p,\ell} \sigma_{{\rm abs};p\ell} \ , \quad
 \sigma_{\rm ext} = \sum_{p,\ell} \sigma_{{\rm ext};p\ell}.
\lb{csisum}
\end{equation}
It is easy to verify that in each particular channel one has $\sigma_{{\rm ext};p\ell} =\sigma_{{\rm sca};p\ell} + \sigma_{{\rm abs};p\ell}$.
For sufficiently small spherical particles of radius $a$, the cross sections Eq.~(\rf{csisum}) are often approximated by the very first electric dipole ($\ell=1$, $p=E$) term in the familiar Rayleigh limit,
\begin{equation}
T_{E1} \to T_{E1;R}= \fr{2ix^3}{3} \frac{\veps-1}{\veps+2} \qquad (x\ll 1),
\label{sphrp}
\end{equation}
where $x=2\pi a/\lambda$, with $\lambda$ being the wavelength in the host medium, is the familiar size parameter~\cite{SI_Bohren1998}.
This enables one an intuitive understanding of small nanoparticles~\cite{SI_Bohren1998}.

Obviously, the above Eqs.~(\rf{sgsc})-(\rf{sgtot}) require a {\em nonabsorbing} host.
The traditional scattering theory neglects the host dissipation and gain~\ct{SI_Newton1982,SI_Bohren1998}, because those cases imply either vanishing or infinite scattering wave at the spatial infinity.
Once $k$ is a complex number, Eqs.~(\rf{sgsc})-(\rf{sgtot}) cannot be straightforwardly extended for $k''\ne 0$, because the expressions yield cross sections as {\em complex} quantities.  

In order to arrive at the extinction cross section in an absorbing host, Bohren and Gilra~\cite{SI_Bohren1979} concluded that it is necessary to move the $1/k^2$ prefactor on the rhs of Eq.~(\rf{sgtot}) under the $\mb{Re}$ sign (see Eq.~(11) in ref.~\citenum{SI_Bohren1979}). Another their peculiar observation was that $\sigma_{\rm ext} \neq \sigma_{\rm sca} + \sigma_{\rm abs}$~\cite{SI_Bohren1979}.
Nearly two decades later, Bohren and Gilra result~\cite{SI_Bohren1979} was in 2018 corrected by Eq.~(45) of Mishchenko and Yang~\cite{SI_Mishchenko2018} in that only the $1/k$-prefactor on the rhs of Eq.~(\rf{sgtot}) goes under the $\mb{Re}$ sign, whereas the second $1/k$-prefactor remains before the real sign, but not as such, because it is, in general, a complex quantity, but as $1/k'$, where $k'$ stands for the {\em real} part of $k$.

% Bibliography
\bibliographystyle{unsrt} 
\bibliography{mlwa_abs_suppl}